\documentclass[fleqn,10pt]{wlscirep}
\usepackage{dcolumn,bm,wrapfig,chngcntr,tocloft,amsmath,amssymb}
\usepackage[utf8]{inputenc}
\usepackage[caption=false]{subfig}
\usepackage[T1]{fontenc} \newcommand{\ello}{\ell_0}
\newcommand{\re}{\text{e}}
\newcommand{\ri}{\text{i}}
\newcommand{\emia}{\re^{-\ri\alpha}}
\newcommand{\epia}{\re^{\ri\alpha}} \newcommand{\muott}{(\mu_1,\mu_2,\mu_3)}
\newcommand{\epitwoa}{\re^{2\ri\alpha}}
\newcommand{\epithreea}{\re^{3\ri\alpha}} \newcommand{\dd}{\text{d}}
\newcommand{\uth}{$^{\text{th}}$\ }
\title{Hydrodynamic object identification with artificial neural models}
\author[1]{Sreetej Lakkam} \author[1]{Balamurali B T} \author[1,*]{Roland
  Bouffanais} \affil[1]{Singapore University of Technology and Design, 8
  Somapah Road, Singapore 487372, Singapore} \affil[*]{bouffanais@sutd.edu.sg}

\begin{abstract}
  The lateral-line system that has evolved in many aquatic animals enables
  them to navigate murky fluid environments, locate and discriminate
  obstacles.  Here, we present a data-driven model that uses artificial neural
  networks to process flow data originating from a stationary sensor array
  located away from an obstacle placed in a potential flow.  The ability of
  neural networks to estimate complex underlying relationships between
  parameters, in the absence of any explicit mathematical description, is
  first assessed with two basic potential flow problems: single source/sink
  identification and doublet detection. Subsequently, we address the inverse
  problem of identifying an obstacle shape from distant measures of the
  pressure or velocity field.  Using the analytical solution to the forward
  problem, very large training data sets are generated, allowing us to obtain
  the synaptic weights by means of a gradient-descent based optimization. The
  resulting neural network exhibits remarkable effectiveness in predicting
  unknown obstacle shapes, especially at relatively large distances for which
  classical linear regression models are completely ineffectual.  These
  results have far-reaching implications for the design and development of
  artificial passive hydrodynamic sensing technology.
\end{abstract}
\begin{document}

\flushbottom
\maketitle
\thispagestyle{empty}

\section*{Introduction}

\label{S:1}

Obstacle detection and identification is instrumental to fish and many
amphibians evolving in aquatic environments. This unique capability of
detecting the surrounding environment from pressure and velocity sensing is
best exemplified by some Mexican blind cave fish that are capable of mapping
cluttered areas only by means of hydrodynamic sensing using their lateral-line
system (LLS)~\cite{campenhausen81:_detec_of_station_objec_by,%
  perera04:_spatial_param_encod_in_spatial,perera05:_later_in_non_visual_sensor,chambers2014fish}.

For engineered vehicles, the ability to identify obstacles and fluid dynamic
conditions of interest can enable efficient path planning while exploiting
favorable environment dynamics~\cite{triantafyllou2016biomimetic}. Sonar,
acoustic Doppler current profiler, and LiDAR are traditionally used to sense
and track environmental features (e.g., obstacles, other vehicles,
etc.). However, these sensors suffer from blind spots and become inoperative
in highly confined, turbid and murky environments. Moreover, stealth
technologies can render targets of interest invisible. To overcome these
limitations, several recent attempts have been made to design artificial flow
sensors by mimicking the basic principles of the
LLS~\cite{chambers2014fish,liu2007micromachined,kottapalli2014touch,%
  dusek2012development,fernandez2011lateral,Asadnia_2015}. Thanks to rapid
developments in MEMS technology, hydrodynamic sensing could soon become a
critical feature of autonomous surface and underwater vehicles performing
navigation tasks in the marine environment.

Although significant technological advances have been achieved towards the
development of an artificial LLS, a full understanding of the underpinning
neural data-processing occurring in aquatic animals equipped with a LLS is
still lacking. To date, two basic hydrodynamic models of obstacle
identification have been proposed based on two different hierarchical
expansions of the flow
field~\cite{sichert2009hydrodynamic,bouffanais2010hydrodynamic}. Sichert~\textit{et
  al.} introduced a hydrodynamic multipole expansion of the velocity potential
and used a maximum-likelihood estimator of linearized relations to estimate
location and shape of the
obstacle~\cite{sichert2009hydrodynamic}. Bouffanais~\textit{et al.} proposed
an obstacle representation using a conformal mapping approach combined with a
general normalization procedure, revealing the progressive perceptual
character of hydrodynamic imaging in the potential flow
framework~\cite{bouffanais2010hydrodynamic}. This hydrodynamic object shape
representation was subsequently extended to Stokes flow by Bouffanais \&
Yue~\cite{bouffanais10:_hydrod_of_cell_cell_mechan}. Given the nonlinear
character of the relationship between local flow data and obstacle shape
parameter, an unscented K\'alm\'an filter---a robust dynamic probabilistic
signal filtering technique for highly nonlinear systems---is used to process
the pressure data gathered by a moving sensor. However, with both
representations, the classical data processing approaches considered---maximum
likelihood estimator and dynamic filtering respectively---exhibit serious
limitations in terms of their ability to extract the features characterizing
the obstacle, and that, even at relatively small distances away from it. This
represents a serious impediment to the actual development of effective
artificial LLS. This issue should come as no surprise given that in the
natural world, the effectiveness of this unique sensory system critically
depends on complex neural data-processing within the central nervous system of
the organism.

Over the past five years, advanced machine learning techniques, and deep
neural networks in particular, have become the prime choice for most problems
categorized as intractable by classical data-mining approaches such as linear
regression or decision tree classifiers. Indeed, artificial neural networks
(ANN) have repeatedly demonstrated their superior performance on a wide
variety of tasks including speech recognition, natural language processing,
image classification, etc. It is worth stressing that this breakthrough in
artificial intelligence is largely due to the ability of performing
unprecedented training of these artificial networks owing to a combination of
high processing power and availability of excessively large training data
sets. In fluid mechanics, ANN have been applied to the study of turbulent
flows, as a data-mining tool to build predictive models associated with direct
numerical simulations. Specifically, these predictive models have been used to
obtain correction factors in turbulent production
terms~\cite{zhang2015machine} or to estimate flow
uncertainties~\cite{ling2015evaluation}. ANN have also been employed to
improve and facilitate turbulence
modeling~\cite{tracey2013application,tracey2015machine,ling2016reynolds}. Deep
neural networks have enabled a novel analysis of turbulent flow fields by
banking on the higher dimensional data associated with rotational and
intermittent turbulent eddies~\cite{kutz2017deep}, thereby revealing that ANN
are significantly more accurate than conventional Reynolds-averaged
Navier--Stokes models. Very recently, ANN have also been used for solving an
engineering problem of obstruction detection in flow
pipes~\cite{carrillo2017recognition}.

Here, we develop an advanced data-driven model based on ANN to address the
shortcomings of previously used data-processing techniques for the problem of
hydrodynamic object recognition, within the potential flow context. We first
study and quantify the ability of ANN in localizing and characterizing
classical potential flow singularities, such as source/sink and doublet. This
allows us to evaluate the influence of the design of the sensor array on the
overall effectiveness of the ANN. As a second step, we consider the
challenging problem of obstacle identification using the shape representation
proposed by Bouffanais~\textit{et
  al.}~\cite{bouffanais2010hydrodynamic}. Remarkably, the ANN are capable of
accurately identifying the shape parameters characterizing the obstacle,
including at relative large distances away from the latter. Moreover, our
ANN-based approach outperforms classical linear regression models, which are
shown to be completely ineffectual for the range of cases considered.

\section*{Methods}

\subsection*{General considerations}

The problem of obstacle shape identification is intrinsically ill-posed given
the lack of an explicit relationship between the local, static, and finite set
of sensed data, on the one hand, and the shape of the obstacle, or the
characteristics of the potential flow singularities considered on the other
hand. For instance, it is expected that the shape parameters become highly
sensitive to minute variations in the sensed data as those are extracted from
increasingly large distances away from the obstacle. Machine learning
techniques are specifically sought after here since they are known to
effectively uncover such unknown relationships between system parameters, even
in the presence of high sensitivity to input data as is the case here.

The definition and mathematical formulation of the studied problem follow the
ones presented in our previous work~\cite{bouffanais2010hydrodynamic}, albeit
with some notable differences. For instance, here we consider that static
velocity or pressure sensing is available, whereas a moving pressure sensor
was considered in Ref.~\cite{bouffanais2010hydrodynamic}. For the sake of
consistency and clarity, some essential elements of
Ref.~\cite{bouffanais2010hydrodynamic} are repeated here, albeit limited to
the necessary level of details.

\begin{figure}[!htbp]
  \centering \subfloat[]{ \includegraphics[width=
    0.55\columnwidth]{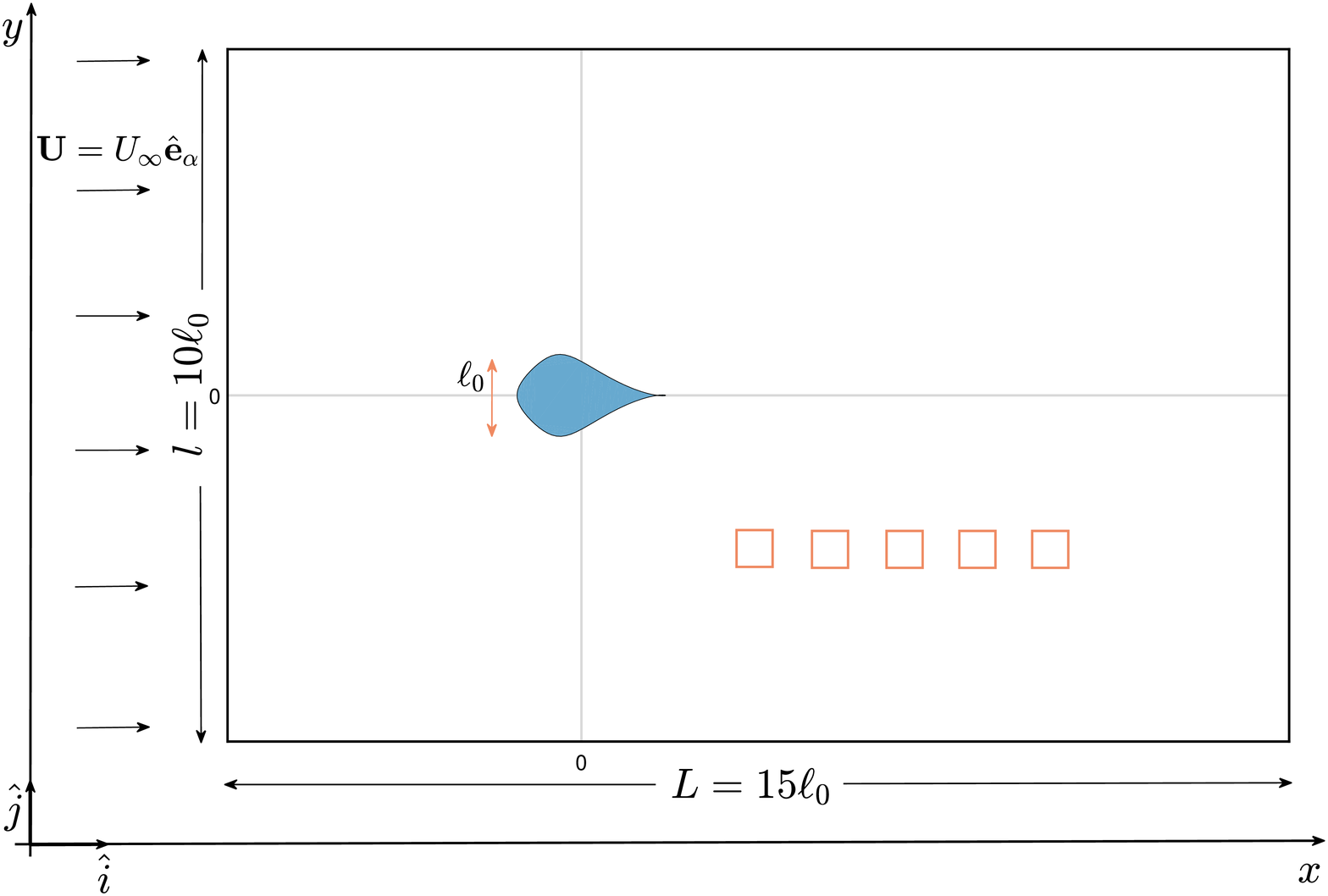} \label{fig:Computational Domain}}
  \subfloat[]{ \includegraphics[width=
    0.35\textwidth]{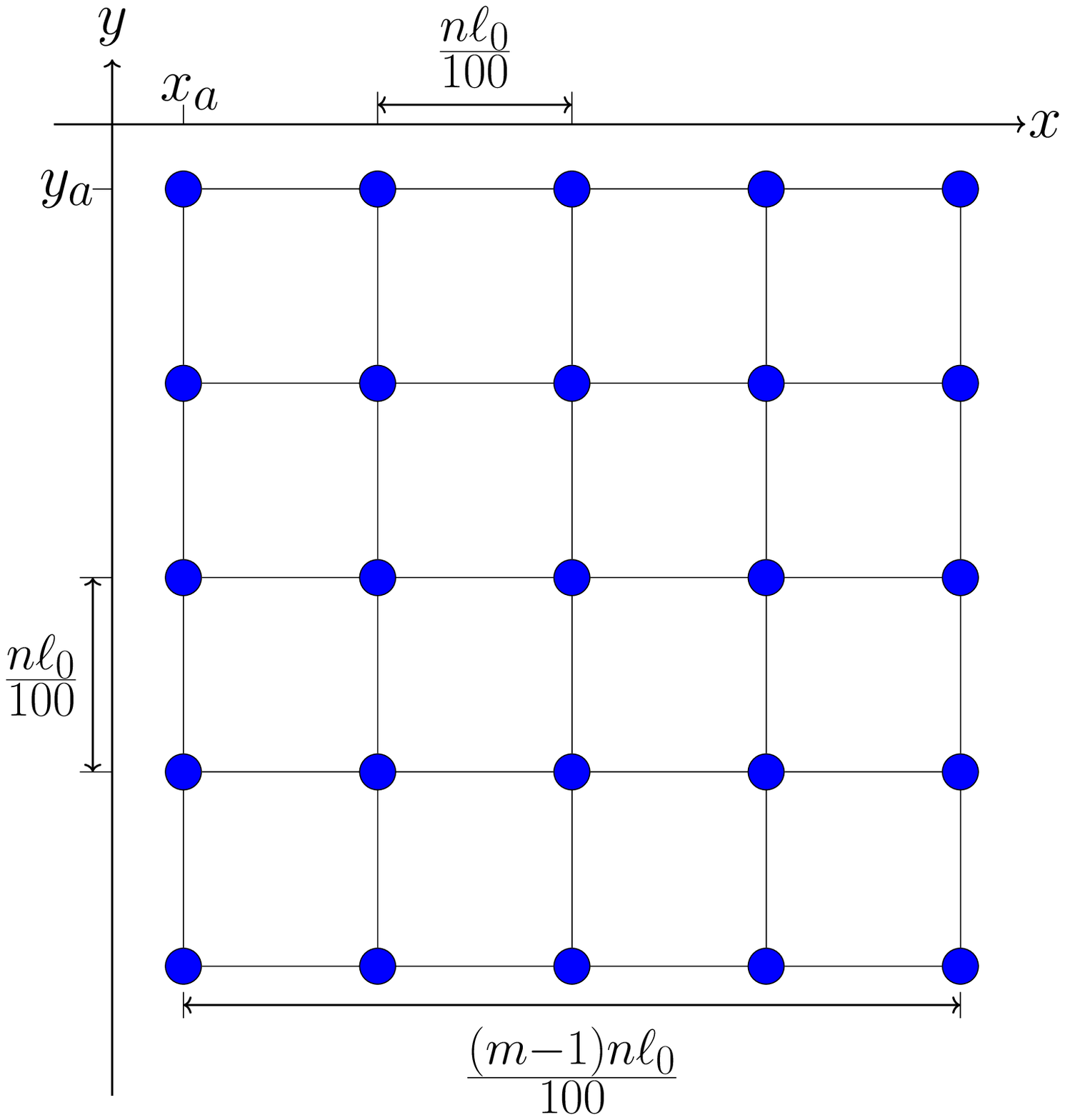} \label{Data stencil schematic}}
  \caption{(a) Schematic diagram of the computational domain, with the object
    to be identified shown in blue (characterized by shape coefficients
    $\mu_1=1/3$, $\mu_2=1/6$, and $\mu_3=1/12$, and the conformal diameter
    $\ello=2R_0$). A few examples of possible locations for the sensor array
    are shown in red. The extracted data is used for training the ANNs. (b)
    Schematic representation of a $(m=5,n)$ sensor array with $25$ sensing
    units uniformly separated by $n\ello/100$. The position of the sensor
    array corresponds to the position $(x_a,y_a)$ of the top-leftmost sensing
    unit.}

\end{figure}

Pressure or velocity sensing are independently available over a static
grid-like sensor array located at some adjustable distance away from the
obstacle to be identified (see Fig.~\ref{fig:Computational Domain}). Note that
the mechanosensory LLS is composed of two sensing units (superficial and canal
neuromasts) giving animals access to both velocity and pressure
sensing~\cite{coombs2003information}. Specifically, the inverse problem
consists in estimating as accurately as possible the shape of an obstacle from
a {\em static} sample of flow data. It is worth highlighting that this inverse
problem, although closely related to the one in
Ref.~\cite{bouffanais2010hydrodynamic}, is considerably more challenging given
the small and finite size of the sample of input data considered here to solve
this problem. For instance, we show in what follows that a few tens of
closely-spaced sensors are amply sufficient to the effectiveness of our
data-driven approach. In comparison, sensed flow data are continuously added
to the input sample, without any restriction, to achieve accurate hydrodynamic
imaging in Ref.~\cite{bouffanais2010hydrodynamic}.

The shape of an obstacle is mathematically related to the sensed data by means
of a nonlinear functional that depends on the particular details of the fluid
flow around the obstacle. It is worth noting, that the problem is further
complexified by the nonlocal relationship between velocity and pressure fields
with incompressible flows. Without lack of generality, we consider
two-dimensional (2D) potential flows, which is a good representation of
three-dimensional shallow-waters with vertical column-like obstacles of
unknown shape used by S. Coombs~\cite{coombs08} to study the LLS-based mapping
behavior of blind cave fish. Moreover, limiting our analysis to relatively
slow fluid flows (maximum speed of the order of one obstacle size $\ello$ per
second, see Fig.~\ref{fig:Computational Domain}), viscous effects and vortex
shedding can be neglected~\cite{triantafyllou00:_hydrod}.

\subsection*{Identification of elementary potential flows}
\label{sec:singularities}

Prior to introducing the complex problem of object shape identification at the
core of this work, we first consider the problem of identification of basic
singularities, which constitute the building blocks of the potential flow
theory, specifically source/sink and doublet. The small number of parameters
involved in fully characterizing these singularities make them good candidates
for ANN-based recognition. The objective is to use these canonical flows to
thoroughly analyze and assess the effectiveness of our data-driven approach
(see Results \& Discussion) in terms of design of sensor and varying
number/nature of parameters to be identified by the ANN.

First, we consider the canonical source/sink flow associated with a velocity
potential
\begin{equation}
  \phi = \frac{q\sigma}{2\pi} \ln r= \frac{q\sigma}{2\pi} \ln  \sqrt{(x-X_s)^2 +(y-Y_s)^2 }, \label{eq:source}
\end{equation}
where $\sigma=\pm 1$ for a source (resp. sink), $q$ is the strength of the
flow (i.e. flowrate per unit depth in 2D), and $(X_s,Y_s)$ is the location of
the singularity in the domain $\overline{\mathcal{D}}$ of interest. Given the
irrotational character of potential flows, the velocity field is classically
given by $\mathbf{v}=u(x,y)\mathbf{i}+v(x,y)\mathbf{j}={\bm \nabla} \phi$ in
2D. To further simplify the problem, we consider $q=2\pi$. To fully
characterize this potential flow, one has to estimate the triplet
$(\sigma,X_s,Y_s)\in \{-1,+1\}\times \overline{\mathcal{D}}$.

The dataset used to train our artificial neural model is obtained by following
these steps: (i) generate a large number of samples for the triplet
$(\sigma,X_s,Y_s)$, with random binary values for $\sigma$ and quasi-random
low-discrepancy Sobol sequences for both $X_s$ and $Y_s$, (ii) for each sample
source/sink $(\sigma,X_s,Y_s)$ generated, the velocity potential $\phi$ is
obtained from Eq.~\eqref{eq:source}, with the $x$-component of the velocity
field given by $u(x,y)=\partial \phi/\partial x$, and (iii) from the field
$u(x,y)$ for each sample $(\sigma,X_s,Y_s)$, one can calculate the sensed
values $u(x_i,y_i)$ at each sensing unit (located at $(x_i,y_i)$) of a given
array.

The second canonical potential flow considered is the doublet flow, which
corresponds to the linear superposition of two closely distant source and
sink. In the far-field approximation, the velocity potential is
\begin{equation}
  \phi = \frac{\kappa}{2\pi}\frac{\cos(\theta - \alpha)}{r}=\frac{\kappa}{2\pi}\frac{\cos(\theta - \alpha)}{\sqrt{(x-X_c)^2 +(y-Y_c)^2 }},\label{eq:doublet}
\end{equation}
where $\kappa = q\ell$, with $\ell$ being the distance between source and
sink, both of strength $q$ (see Eq.~\eqref{eq:source}). The center of the
doublet is located at $(X_c,Y_c)$, and at equal distance of source and
sink. The angle $\alpha$ is the doublet orientation, measured from the
$x$-axis corresponding to $\theta=0$ in domain $\overline{\mathcal{D}}$. To
further simplify the problem, we consider $\kappa=2\pi$. To fully characterize
this doublet flow, one has to estimate the triplet $(\alpha,X_c,Y_c)\in
(-\pi,\pi]\times \overline{\mathcal{D}}$. It appears therefore clearly that
the increased complexity in identifying this doublet flow as compared to a
source/sink one, comes from the estimation of $\alpha$ as opposed to the
simpler binary classification for $\sigma$.

The process to generate the training dataset for this doublet flow is
essentially the same as the one for the source/sink flow, except for the fact
that in step (i), one has to generate a large number of samples for the
triplet $(\alpha,X_c,Y_c)$, with quasi-random low-discrepancy Sobol sequences
for $\alpha$, $X_c$ and $Y_c$. The other steps being identical.

\subsection*{Object shape representation}
\label{sec:shape}

We now consider the problem of identifying of the shape of a single obstacle
placed in a potential flow. The set of 2D located shapes is infinite
dimensional, thereby stressing the inherent challenges associated with such
shape classification process~\cite{sharon20062d}. Before proceeding, we note
that our work is aimed at classifying shapes while considering the size and
location of the obstacle as known quantities. This assumption admits two
justifications. First, Burt de
Perera~\cite{perera04:_spatial_param_encod_in_spatial} has proved that blind
cave fish encode independently in its spatial map obstacle size and location
on the one hand, and obstacle shape on the other hand. Second, our previous
work~\cite{bouffanais2010hydrodynamic} proved that obstacle size $R_0$ and
location $a_0$ (using the exact same notations as the ones previously
introduced) can relatively easily be determined, even at large distances away
from the object.

The powerful and compact shape representation introduced in
Ref.~\cite{bouffanais2010hydrodynamic} rests upon a conformal mapping
technique of the fluid domain $\mathcal{D}$ exterior to the 2D curve limiting
the body of the obstacle $\mathcal{O}$, in the complex $z$-space with
$\overline{\mathcal{D}}=\mathcal{D}\cup \mathcal{O}$. Without loss of
generality, we assume that the obstacle to be identified is centered---i.e.,
the conformal center is $a_0=0$---and has a unit size---i.e., the conformal
radius is unity $R_0=1$ and typical object size $\ello=2R_0=2$. As a
consequence, the inverse exterior Riemann mapping $z=f^{-1}(\xi)=h(\xi)$ from
the exterior of the unit disc $|\xi|>1$ onto $\mathcal{D}$ takes the form of
the following Laurent series, with a simple pole at infinity:
\begin{equation}
  \label{eq:LS}
  z=h(\xi) = \xi +
  \frac{\mu_1}{\xi}+\frac{\mu_2}{\xi^2}+\cdots ,\qquad |\xi|>1.
\end{equation}
The key idea behind the use of a Laurent series is that the shape information
is encoded into the discrete set of complex coefficients $\{\mu_k\}_{k\geq1}$,
also referred to as the shape `fingerprint'~\cite{sharon20062d}. Note that the
univalent character of the analytic function $h$ imposes the following
constraints on these coefficients:
\begin{equation}
  |\mu_k| \leq \frac{1}{\sqrt{k}}, \qquad k\geq 1.
  \label{eq:muk-bound}
\end{equation}
As detailed in Ref.~\cite{bouffanais2010hydrodynamic}, each term in $1/\xi^k$
in Eq.~\eqref{eq:LS} is associated with a specific geometric polygonal
perturbation of the unit circle $|\xi|=1$---specifically a $(k+1)$-gonal
perturbation. For instance, the term in $1/\xi^2$ corresponds to an
equilateral triangle perturbation, while the one in $1/\xi^3$ is of the square
type.

Beyond the above mathematical considerations, it is essential to keep in mind
that the hierarchical distribution of high-order terms in $h$ is directly
associated with the progressive perceptual discrimination of the obstacle
shape from a distance. Indeed, at large distance from the obstacle, i.e. for
$|\xi| \gg 1$, only a limited number of terms in $1/\xi^k$ in the asymptotic
expansion of $h$ are sufficient to provide an accurate approximate shape
representation.

We now consider the forward problem associated with any 2D obstacle placed in
a uniform external flow $\mathbf{U}=U_\infty \hat{\mathbf{e}}_\alpha$ making
an angle $\alpha$ with the $x$-axis (see Fig.~\ref{fig:Computational
  Domain}). The complex potential $w$ is obtained through conformal
mapping~\cite{milne1996theoretical}
\begin{equation}\label{eq:w} w=\phi+\ri \psi=U_\infty \left[ \xi \emia +
    \frac{\epia}{\xi}\right],\qquad |\xi|>1,
\end{equation}
where $\psi$ is the streamfunction, and $\xi=f(z)$ is the direct exterior
mapping that is uniquely and completely characterized by the shape
coefficients $\{\mu_k\}_{k\geq1}$. The expression of $w$ in the complex
$z$-space is derived using a series inversion of $z=h(\xi)$ (see
Eq.~\eqref{eq:LS}), up to a given order. Without loss of generality, we limit
our shape fingerprint to cases such that $k\leq 3$, and our series inversion
is expanded up to third order terms in $1/z$:
\begin{equation}
  \label{eq:w2}
  w =U_\infty \left[
    z \emia+\frac{\varpi_1\epia}{z}+\frac{\varpi_2\epitwoa}{z^2}+\frac{\varpi_3\epithreea}{z^3} \right]
  +\mathcal{O}\left( \frac{1}{z^4}\right),
\end{equation}
with $ \varpi_1 = 1-\mu_1\re^{-2\ri\alpha}$, $\varpi_2 =
-\mu_2\re^{-3\ri\alpha}$, and $\varpi_3 =
\mu_1\re^{-2\ri\alpha}-(\mu_1^2+\mu_3)\re^{-4\ri\alpha}$. Note that
Eq.~\eqref{eq:w2} for $w$ ceases to be valid at too short distances away from
the considered obstacle. However, this limitation has absolutely no impact on
the present study given that both training and classification occur at
sufficiently large distances away from the obstacle. The complex flow velocity
is classically obtained by
\begin{equation}
  u+\ri v =\frac{\dd w}{\dd z},\label{eq:uiv}
\end{equation}
giving access to the $x$- and $y$-component of the velocity field
$\mathbf{v}$. Lastly, the normalized pressure is cast as
\begin{equation}
  p(z)=-\frac{1}{2} \left| \frac{\dd w}{\dd z}\right|^2.\label{eq:p}
\end{equation}
Equations~\eqref{eq:uiv} \&~\eqref{eq:p} constitute the analytical solution to
the forward problem associated with our inverse problem of shape
identification.

Although the process of generating the training dataset for this forward
problem is built on similar steps as those detailed in previous sections, it
is detailed in the next section due to its technicality.

\subsection*{Data generation}
\label{sec:training}

All lengths are considered in units of the object size $\ello$. For the
source/sink/doublet identification cases, the domain is a square one of size
$10\ello\times 10\ello$, uniformly discretized with a grid of $1000\times
1000$ points. For the object shape identification problem, we consider a
rectangular domain of size $15\ello\times 10\ello$, uniformly discretized with
a grid of $1500\times 1000$ points along the $x$- and $y$-direction
respectively. Moreover, all objects are considered to have the same size
$\ello$, are centered at the origin, and only differ in their shape fully
characterized by the set of first three coefficients $\{\mu_1, \mu_2, \mu_3\}$
(see Fig.~\ref{fig:Computational Domain}). By construction, the object is
symmetric with respect to the $x$-axis, but this symmetry in the problem can
be removed by changing the incidence $\alpha$ of the upstream uniform flow
$\mathbf{U}=U_\infty \hat{\mathbf{e}}_\alpha$.

The dataset used to train and test our artificial neural network model is
obtained by generating a large number of samples ($N=10^5$ in total, with
$90\%$ samples for training and $10\%$ for testing purposes). Each sample
corresponds to a unique obstacle shape fully characterized by the triplet
$\muott$. We obtain the training-cum-testing dataset by generating three
quasi-random, low-discrepancy Sobol sequences of size $N$ for $\mu_1$,
$\mu_2$, and $\mu_3$ while accounting for the individual constraint on each
$\mu_k$ given by Eq.~\eqref{eq:muk-bound}. For each sample, the forward
problem presented is solved analytically, thereby yielding both components of
the velocity field from Eq.~\eqref{eq:uiv}, and the dynamic pressure from
Eq.~\eqref{eq:p}. It is worth stressing that this step---the generation of the
training dataset---is often the limiting one when using deep neural networks
applied to hydrodynamics problems. Indeed, the effectiveness of such methods
critically depends on having access to significantly large datasets, which is
often impractical both experimentally and numerically. The last step is
trivial and consists in calculating the sensed values $u(x_i,y_i)$,
$v(x_i,y_i)$, and $p(x_i,y_i)$ at each sensing unit (located at $(x_i,y_i)$)
of a given array.

\subsection*{Data extraction and feature space projection}

The estimation of location for data extraction in flow domain is paramount for
training the model. Data capturing the whole range of variations in flow field
aids in obtaining accurate relations between the flow velocity vector field
and the shape coefficients. An important element for data extraction is the
design of the sensor, i.e. number of data points and the spacing between the
data points. A detailed analysis is performed to determine an adequate sensor
array design, comparing prediction model performance in identification of
elementary potential flows (see Results \& Discussion).

Upon selecting an appropriate array design and its location for data
extraction, one can start preprocessing the training-cum-testing
dataset. Feature projection methods are known to drastically improve the
performance of classifiers susceptible to the Hughes
phenomenon~\cite{1054102}. For our problem, this approach is particularly
helpful because the intrinsic dimensionality of data is much less than the
number of features dealt with. Specifically, we use the classical linear
principal component analysis~(PCA), which is widely used in modern data
analysis~\cite{DBLP:journals/corr/Shlens14}. Over the past decade, numerous
groups have successfully combined PCA with ANNs to solve a wide range of
problems across multiple
fields~\cite{bucinski2005clinical,sousa2007multiple,liu2007hierarchical,zhang2007artificial}. Essentially,
PCA maps data to a low-dimensional space while maximizing data variance in the
low-dimensional representation. It is important noting that PCA is sensitive
to the relative scaling of the original variables which has important
implications to this work since the three shape coefficients $\muott$ vary in
different intervals given the constraints~\eqref{eq:muk-bound}.

In practice, PCA is a principal axis transformation technique that minimizes
the correlation of variables in a $p$-dimensional space to a $q$-dimensional
subspace with a new basis formed by the linearly uncorrelated principal
components---i.e., eigenvectors. PCA arranges these principal components in a
reducing order of importance in representing input data information. The first
component represents the highest information content while the last one
contains the least. After obtaining the principal components, the new
variables for each sample data are calculated as a linear combination of the
original data variables and the higher-order terms can be neglected for
dimensional reduction~\cite{cao2003comparison}.

The explained variance ratio of a given principal component---ratio between
the variance of that component and the total variance---reveals how much
information can be attributed to each principal component. This is important
as we convert a 9-dimensional input space into a 3-dimensional one. By
estimating the explained variance ratio, we can see that the highest explained
variance for the first principal component is approximately $95.1\%$, while it
is $4.88\%$ for the second principal component contains, and the third
principal component has a very low explained variance of $0.02\%$. Together,
the three principal components represent almost $100\%$ of the information,
while the first two, with $99.98\%$, constitute the bulk of it. These results
are obtained with a square sensor array comprised of 25 sensing points
(optimal $(5,10)$ design identified at a data extraction distance $2.8\ello$
behind the object (leftmost red square in Fig.~\ref{fig:Computational
  Domain}).

The crucial need for preprocessing the dataset by means of PCA is apparent
when plotting the estimated values of the shape coefficients
$\{\tilde{\mu}_k\}_{k=1,2,3}$ against the actual ones $\{\mu_k\}_{k=1,2,3}$
with and without PCA (see Supplementary Material, Note I).

\subsection*{Prediction model}
\label{sec:prediction}

As already mentioned, artificial neural networks are chosen here given their
vastly superior performance in dealing with implicit nonlinear relationships
as compared to traditional methods~\cite{haykin2009neural}. For instance, with
linear regression (LR), nonlinearities have to be known a priori and expressed
explicitly. ANN, however, do not require any a priori knowledge of possible
relations between parameters, and typically infer these relations much more
effectively than multiple regression analyses~\cite{palocsay2004neural}, or
other conventional statistical methods~\cite{scarborough2006neural}. Hence,
ANN provide a nonparametric approach that is free of assumptions and adaptive,
i.e., if new data is available, the model will adapt to
it.~\cite{detienne2003neural}.

Whenever ANN are used, one should always verify that traditional
data-processing methods such as LR are indeed ineffectual. We therefore
evaluate the effectiveness of a classical LR approach in dealing with the
simple elementary doublet flow introduced in Sec.~\ref{sec:singularities}. To
further simplify the problem, we fix the orientation of the doublet to
$\alpha=0$ and the goal is to estimate the location $(X_c,Y_c)$ of the center
of the doublet. Figure~S3 (see Supplementary Material, Note III) clearly shows
that linear regression is absolutely ill-equipped for this basic task. In what
follows, the effectiveness of all prediction models is quantified by means of
the relative error between estimated value $\widetilde{\Sigma}$ and the actual
real value $\Sigma$ of any given property of the system considered:
$\Sigma=X_s$, $Y_s$, $\sigma$, $X_c$, $Y_c$, $\alpha$, $\mu_k$, etc. This
relative error, denoted $\rho$, is defined as
\begin{equation}\label{rho}
  \rho=\frac{\|\widetilde{\Sigma}-\Sigma\|}{\|\Sigma\|},
\end{equation}
and has values in the unit interval.

Given the complete ineffectiveness of LR, artificial neural networks are used
to estimate the object shape coefficients. The ANN architecture considered
here contains three hidden layers with $m=100$, $100$, and $16$ hidden units
$\{H_{i}^{l}\}_{i=1,\dots,m}$ per hidden layer $l$ respectively (see Fig.~S2,
Supplementary Note II). In addition, a rectified linear unit (ReLU) serves as
activation function. In the absence of such an activation function, the neural
networks exhibit poor performance in most of the cases considered, including
with the basic source/sink identification problem (not shown here). The ReLU
activation approach is known to be more effective than other classical
continuous activation functions, e.g. sigmoid or hyperbolic
tangent~\cite{DBLP:journals/corr/abs-1805-08786,DBLP:journals/corr/abs-1710-05941}.

The so-called `Adam' method, which is an extension of the classical
gradient-descent algorithm, is used to stochastically optimize the weights of
the neurons as part of the backpropagation
algorithm~\cite{DBLP:journals/corr/KingmaB14}. Backpropagation acts
iteratively following a two-step sequence consisting of a forward pass that is
followed by a backward one. In the former step, activation value for each
neural unit in the network are calculated from the weights of the adjacently
connected neurons. During the backward pass, weights are corrected based on
the difference between generated output during the forward pass and the
desired output. The ANN architecture described above is implemented by means
of the {\tt \small scikit-learn} machine learning
library~\cite{pedregosa2011scikit}. As already mentioned, for each case
reported in this study, the total number of samples is $N=10^5$, with $90,000$
training samples generated through the process detailed previously, and the
remaining $10,000$ ones used for testing purposes.

\section*{Results \& Discussion}

\subsection*{Influence of the sensor array design}
\label{Stencil analysis}

Unsurprisingly, among the animal taxa afforded with a mechanosensory apparatus
enabling flow detection, there exists a wide range of natural designs: from
the LLS in fish, to patches of whiskers in pinipeds, etc. One is therefore led
to question the influence of the design of the sensor array on the
effectiveness of the hydrodynamic object identification at a distance. It is
worth pointing out that this important question has relevance at the practical
MEMS-device design level, but also at the fundamental data-processing level,
which is the scope of this work.

We consider square sensor arrays fully characterized by a set of integers
$(m,n)$, having $m\times m$ sensing units evenly separated by a spacing of
length $n\ello/100$. The side length of these square arrays is simply
$(m-1)n\ello/100$, expressed in units $\ello$ of the object size (see
Fig.~\ref{Data stencil schematic}). For instance, the $(5,10)$ array comprises
of 25 sensing units evenly separated by $\ello/10$. The position $(x_a,y_a)$
of the sensor array is measured from the origin (conformal center of the
object as in Fig.~\ref{fig:Computational Domain}) and corresponds to the
position of the top-leftmost sensing unit. By varying $m$ and $n$, we can
generate a wide range of sensor arrays of different sizes/resolutions.

\begin{figure}[!htbp]
  \centering \subfloat[$(3,2)$-Sensor array] {\includegraphics[width=
    0.22\columnwidth]{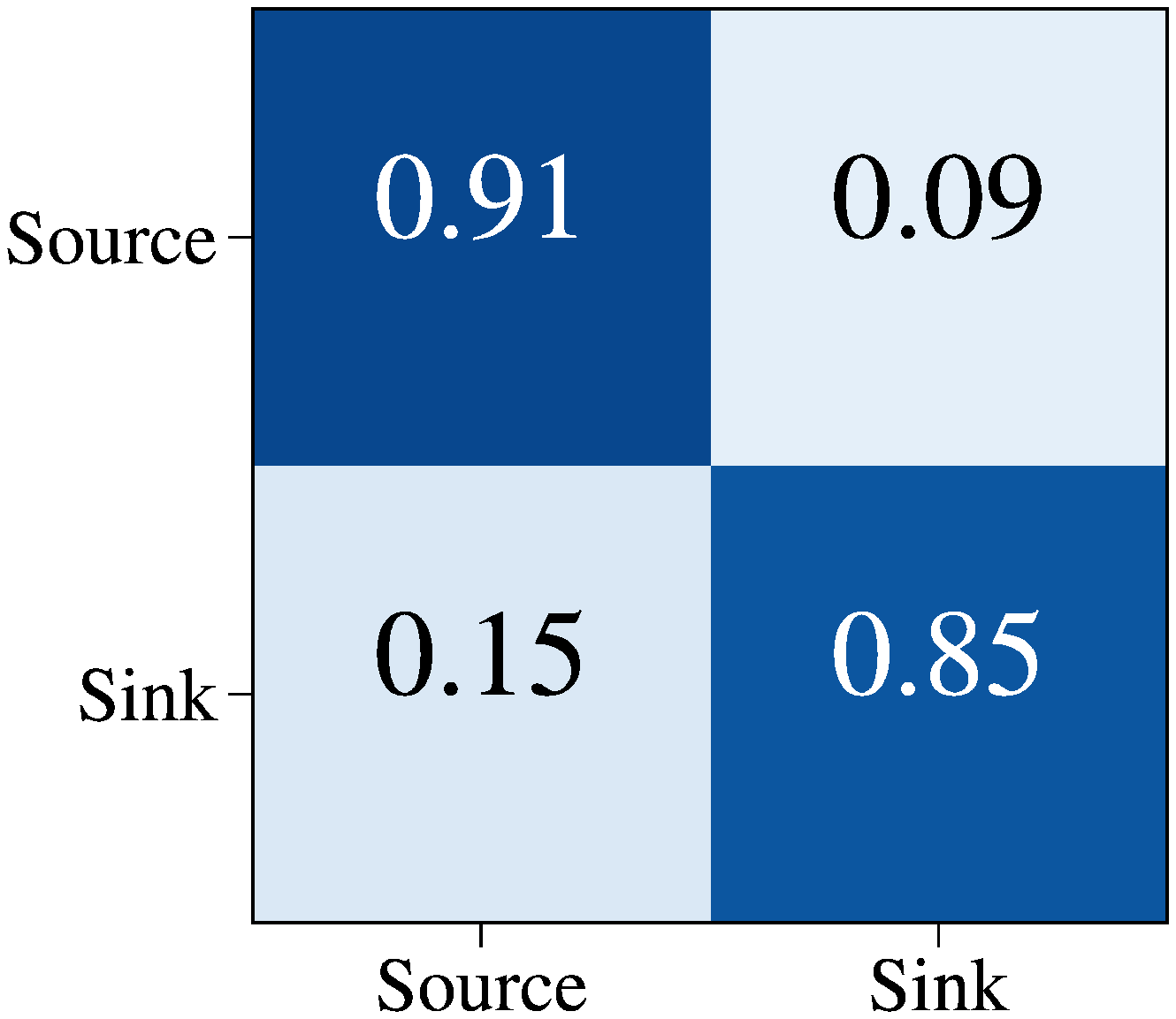}\label{cfm_3point_2spacing}}
  \subfloat[$(3,5)$-Sensor array] {\includegraphics[width=
    0.22\columnwidth]{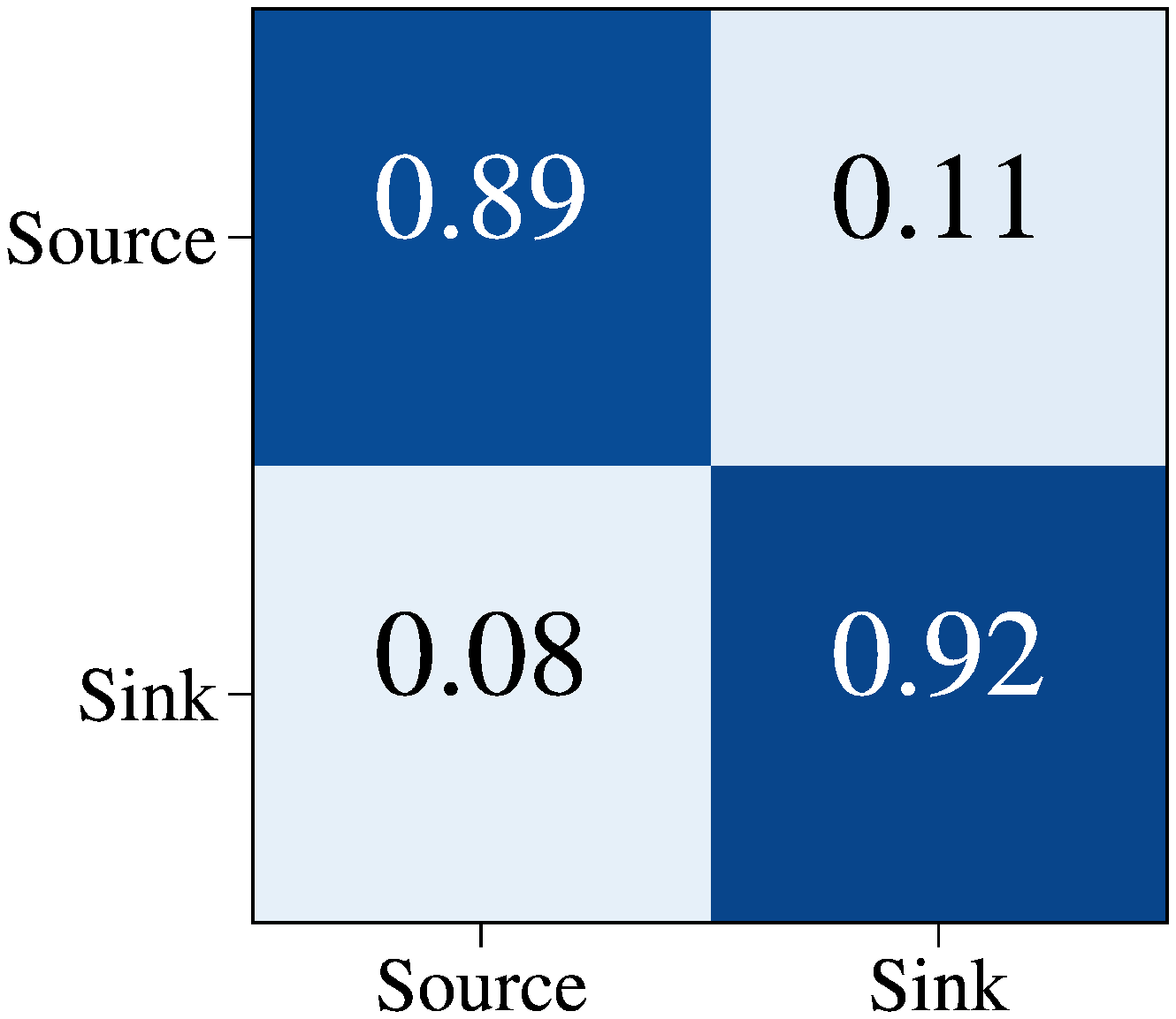}\label{cfm_3point_5spacing}}
  \subfloat[$(3,10)$-Sensor array] {\includegraphics[width=
    0.255\columnwidth]{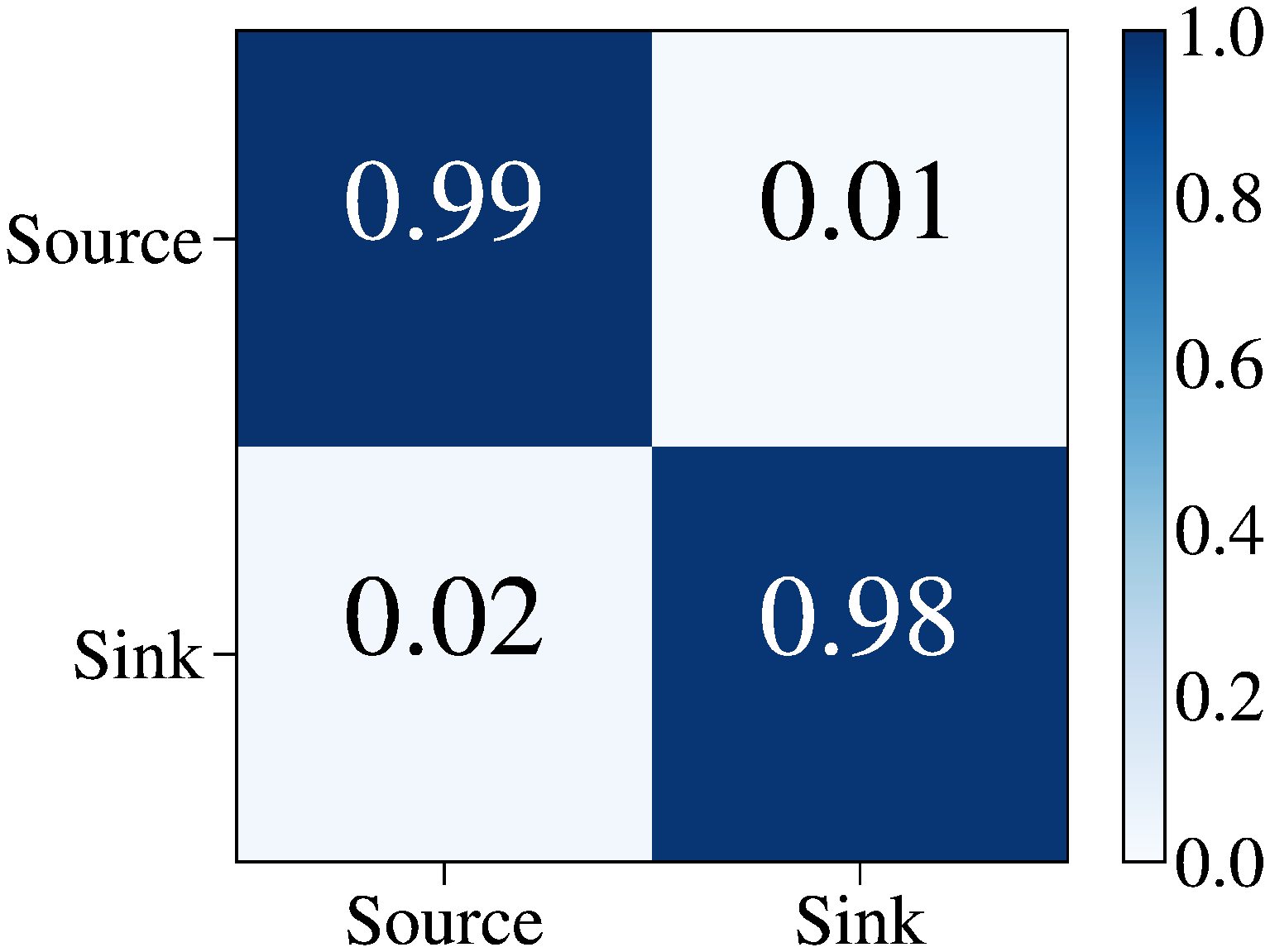}\label{cfm_3point_10spacing}}
  \hspace{5mm} \subfloat[$(5,2)$-Sensor array] {\includegraphics[width=
    0.22\columnwidth]{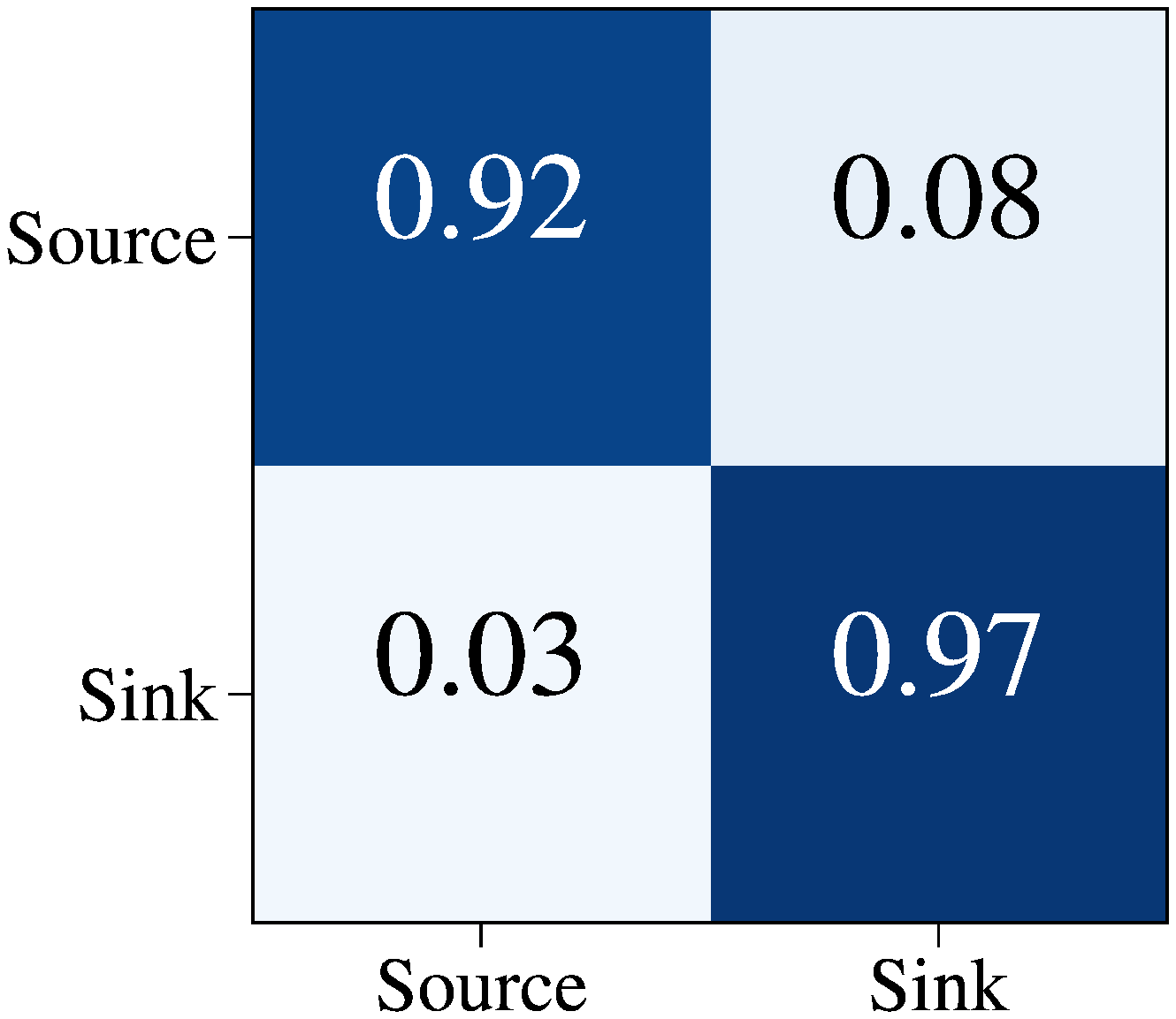}\label{cfm_5point_2spacing}}
  \subfloat[$(5,5)$-Sensor array] {\includegraphics[width=
    0.22\columnwidth]{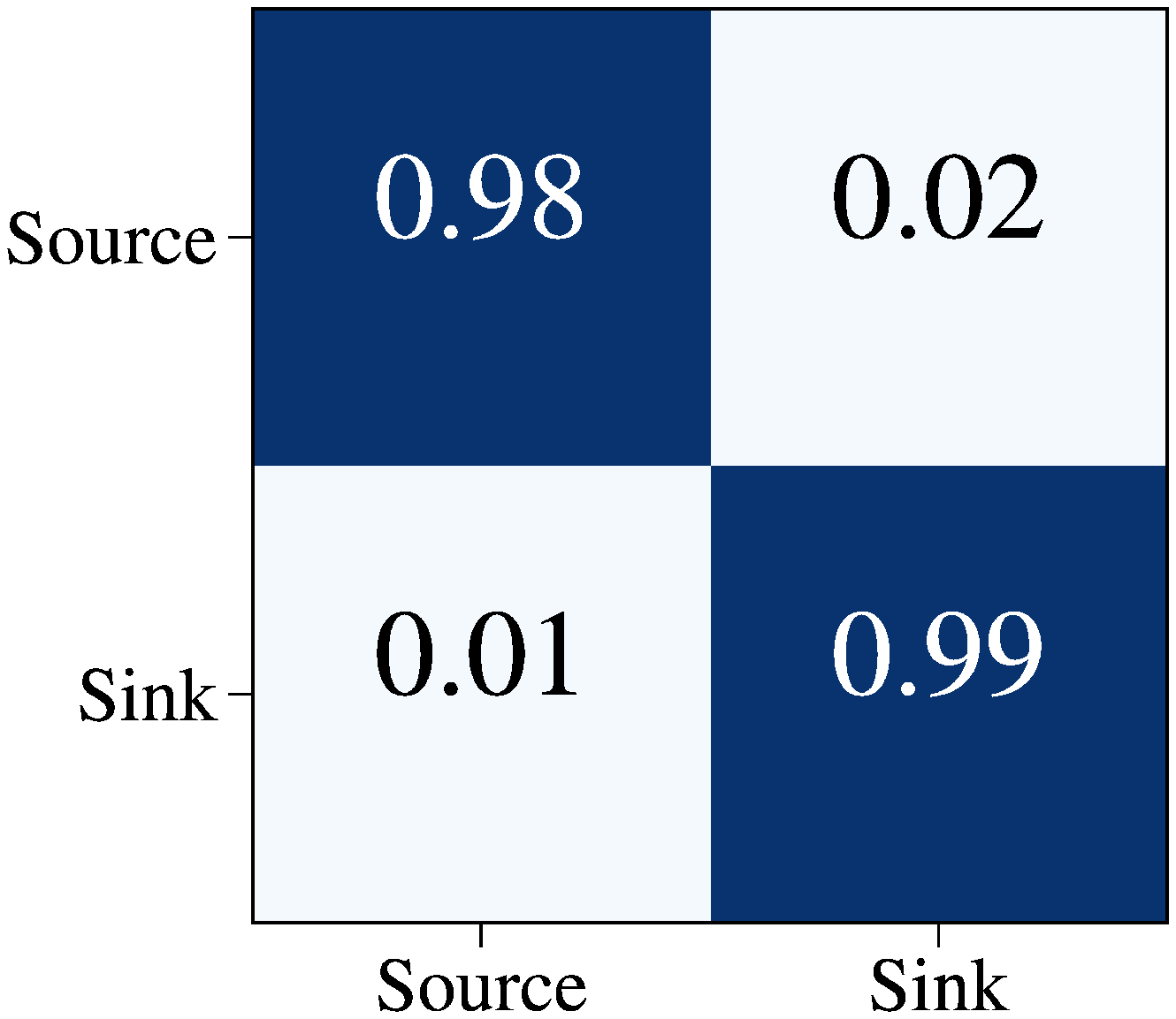}\label{cfm_5point_5spacing}}
  \subfloat[$(5,10)$-Sensor array] {\includegraphics[width=
    0.255\columnwidth]{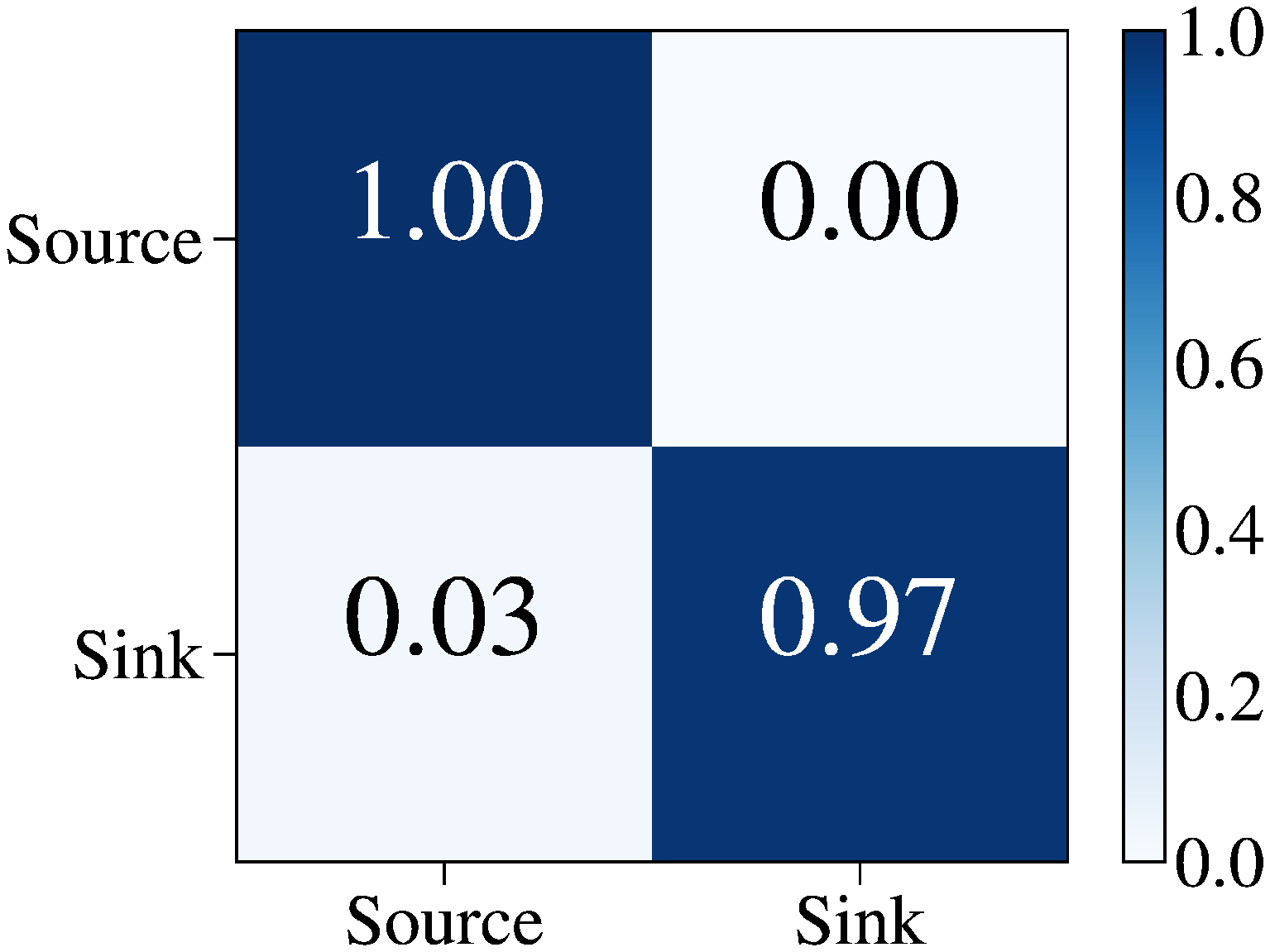}\label{cfm_5point_10spacing}}
  \hspace{5mm} \subfloat[$(7,2)$-Sensor array] {\includegraphics[width=
    0.22\columnwidth]{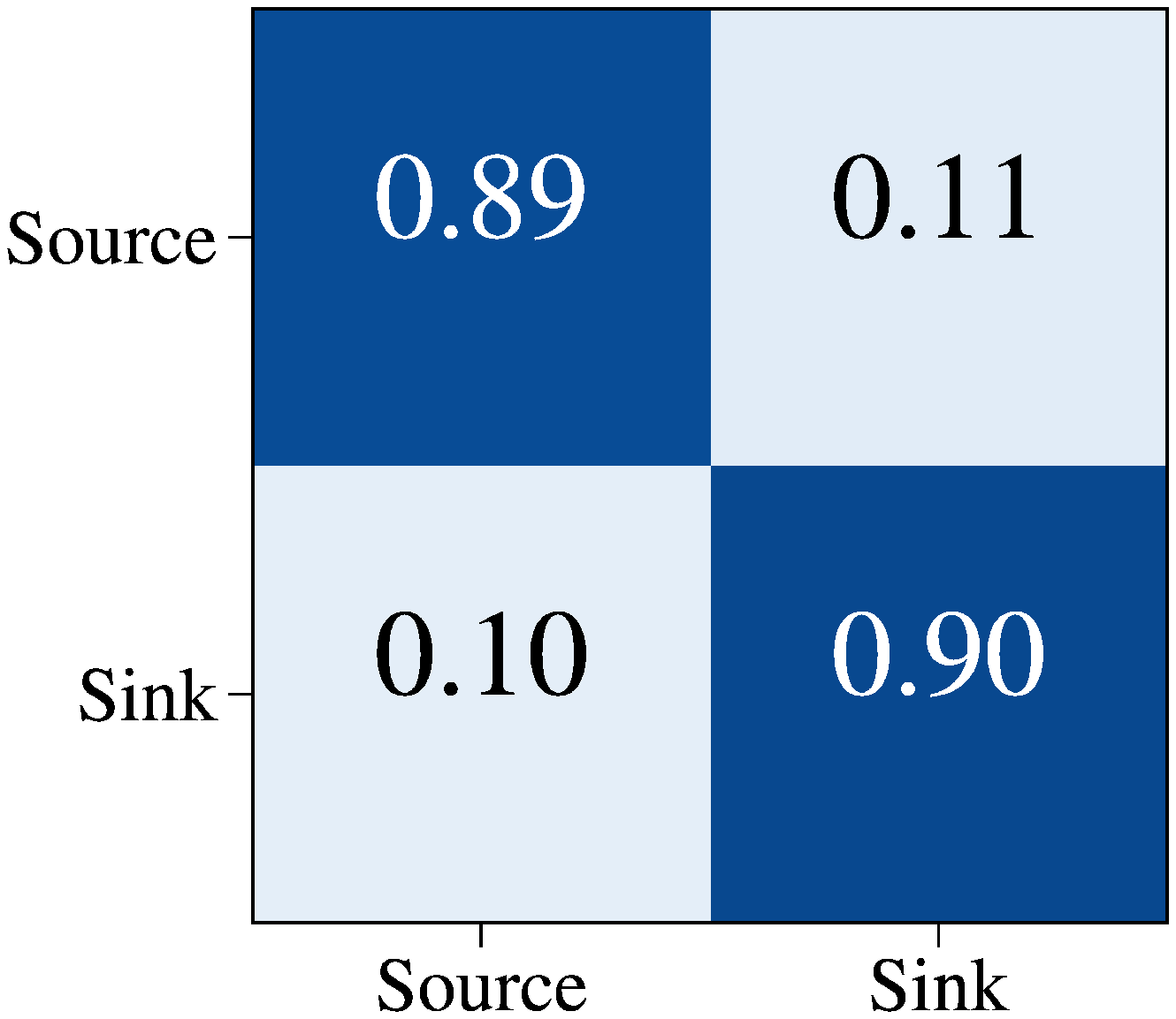}\label{cfm_7point_2spacing}}
  \subfloat[$(7,5)$-Sensor array] {\includegraphics[width=
    0.22\columnwidth]{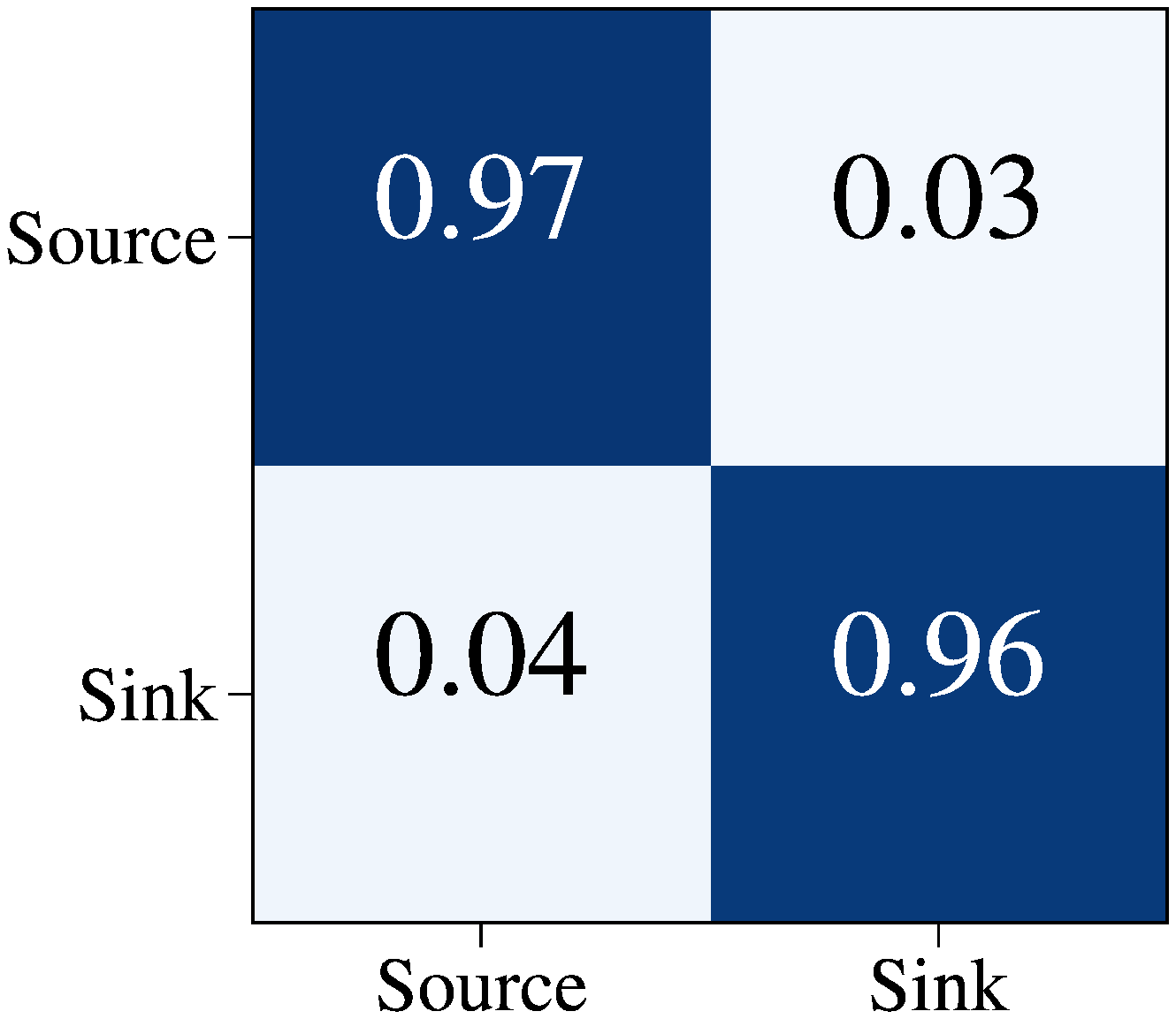}\label{cfm_7point_5spacing}}
  \subfloat[$(7,10)$-Sensor array] {\includegraphics[width=
    0.255\columnwidth]{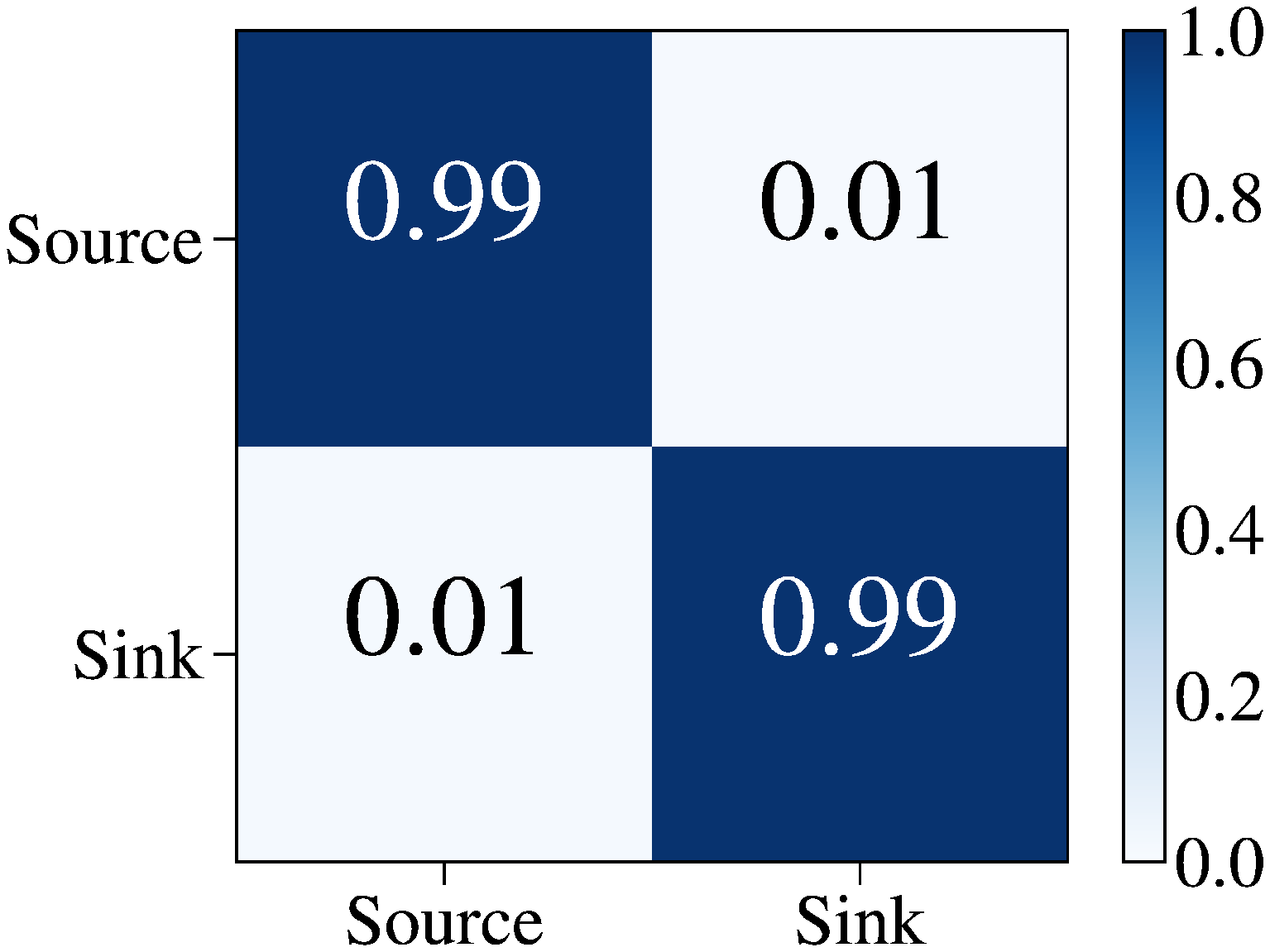}\label{cfm_7point_10spacing}}
  \caption{Confusion matrices for 9 distinct $(m,n)$-sensor arrays performing
    an ANN-based binary classification between source/sink flows. Actual flow
    singularity is on the $x$-axis, while estimated singularity is on the
    $y$-axis. The numerical entries are percentages also shown using a blue
    color map of the unit interval.}
  \label{Classifier accuracy single source}
\end{figure}

We first tackle the ANN-based source/sink flow identification problem (see
Methods) with 9 distinct sensor arrays. It is worth recalling that for this
problem the ANNs are trained with the $x$-component of the velocity field,
$u$, extracted at the $m^2$ sensing points, with the array located at $(x_a =
2\ello, y_a =- 2\ello)$ measured from the origin located at the center of the
square domain of side length $10\ello$. To assess the performance of the
binary classifier distinguishing between source ($\sigma=+1$) and sink
$(\sigma=-1)$, we present the results using confusion
matrices~\cite{haykin2009neural}. These are 2-by-2 tables, that report the
fraction of false positives, false negatives, true positives, and true
negatives, thereby providing an in-depth report of the statistical
classification beyond just the proportion of correct estimations. Essentially,
an effective binary classification leads to a mostly diagonal confusion matrix
(i.e. with high fractions of true positives and true negatives). Off-diagonal
entries represent the percentage of misclassified samples (i.e. false
positives and false negatives). Figure~\ref{Classifier accuracy single source}
shows the confusion matrices corresponding to 9 sensor arrays obtained with
$m^2=9,25,49$ sensing units and $n=2,5,10$. For instance, the $(3,2)$-sensor
array (see Fig.~\ref{cfm_3point_2spacing}) yields on average a 91\% accuracy
in identifying sources, 85\% for sinks, 15\% misclassification of sinks, and
9\% misclassification of sources. Although we notice a systematic improvement
when going from a $3\times 3 $ array to a $5\times 5$ one, further doubling
the number of sensing units ($7\times 7$) leads to a slight reduction in the
effectiveness of the classification, and that for all three values of
$n$. This effect is attributed to the fact that this binary classification is
carried out using data sets obtained from the continuous field
$u(x,u)$. Classification tasks using discrete data sets are usually more
effective.  As expected, increasing $n$ almost always yields a higher
accuracy, which can easily be explained by the increased size of the sensor
array.

Next, we turn to the effectiveness of the regressor in finding the location
$(X_s,Y_s)$ of the flow singularity (source or sink). Figure~\ref{Location
  accuracy single source} shows the relative error (see Eq.~\eqref{rho}) in
the estimated position $\widetilde{X}_s$ and $\widetilde{Y}_s$, with boxplots
representing the the median error in red, and box extremities corresponding to
the 25\uth and 75\uth percentiles of the distribution of error, and whiskers
being at the 10\uth and 90\uth percentiles. Contrary to the previous binary
classification, the ANN-based regressor exhibits a systematic improvement with
increases in both $m$ and $n$. This is particularly clear when observing the
steep narrowing of the whiskers of the distribution of relative error. An
increase in $n$, however, yields a more pronounced improvement of the overall
accuracy of the regressor. This can easily be attributed to the fact that the
associated array design reduces data localization. The difference in
performance between classifier and regressor can be traced to the continuous
nature of the field $u(x,y)$. Again, classifiers are known to be more
effective when learning from discrete data sets.

From Figs.~\ref{Classifier accuracy single source} \& \ref{Location accuracy
  single source}, we can conclude that the sensor array $(m=5,n=10)$ (with 25
sensing units and side length $0.4\ello$) constitutes a satisfactory option
given its high performance as classifier, as well as regressor. For instance,
the array $(m=7,n=10)$ yields marginally better performance although it
comprises of almost twice as many sensing units, and has a sensing area more
than doubled that of $(m=5,n=10)$. In all results that follow, the
$(m=5,n=10)$-sensor array design is systematically used, unless mentioned
otherwise.

\begin{figure}[!htbp]
  \centering \subfloat[$(3,2)$-Sensor array] {\includegraphics[width=
    0.22\columnwidth]{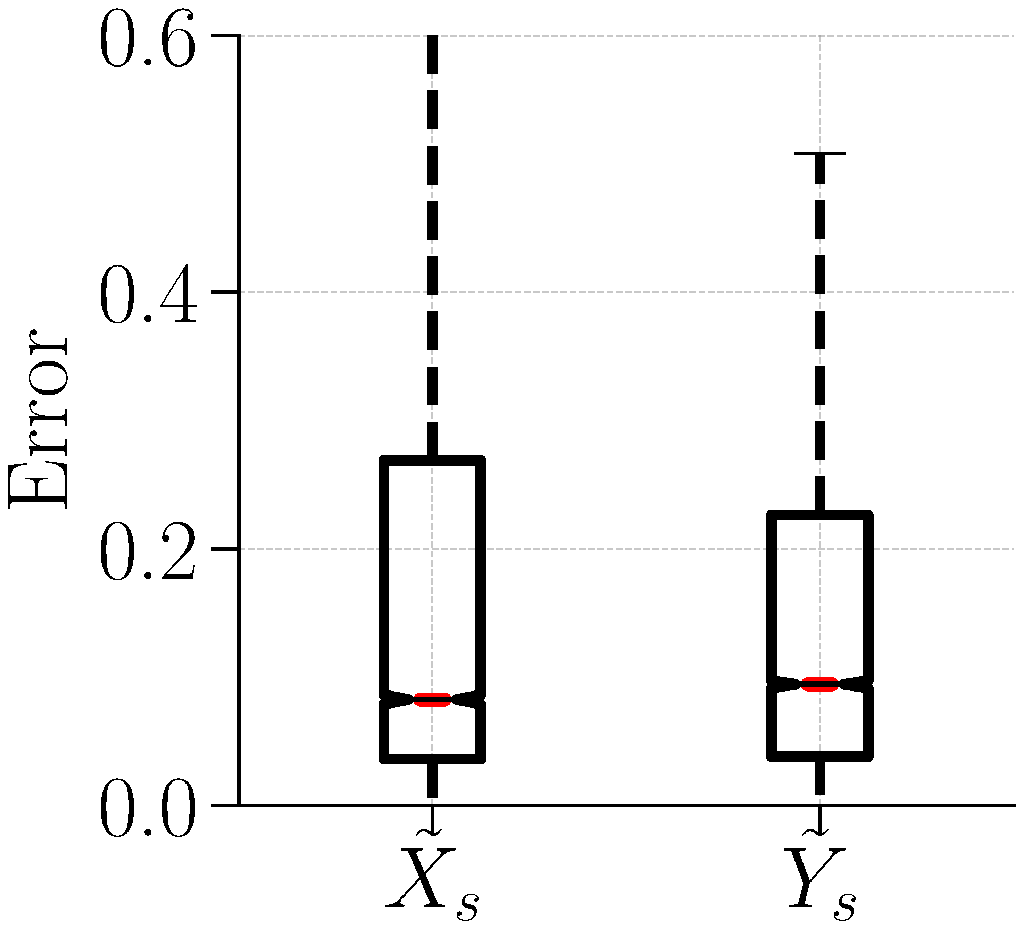}\label{box_plot_3point_2spacing}}
  \subfloat[$(3,5)$-Sensor array] {\includegraphics[width=
    0.22\columnwidth]{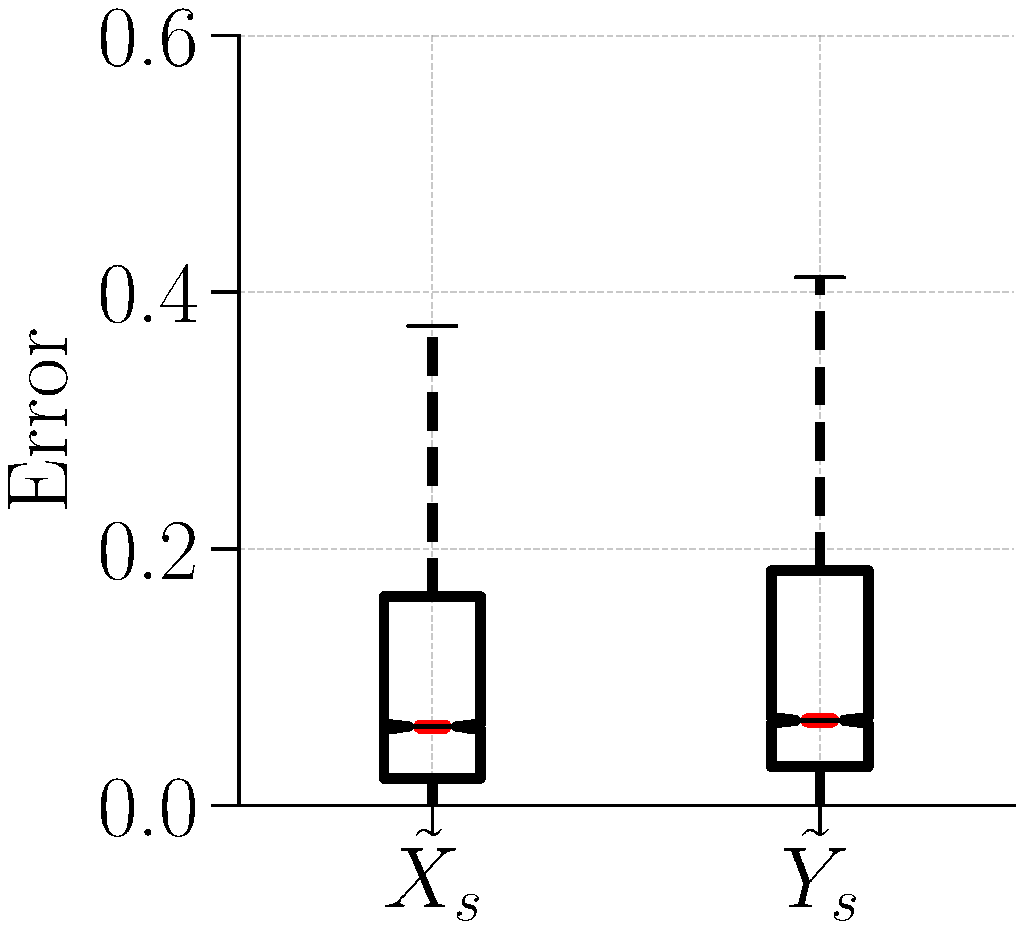}\label{box_plot_3point_5spacing}}
  \subfloat[$(3,10)$-Sensory array] {\includegraphics[width=
    0.22\columnwidth]{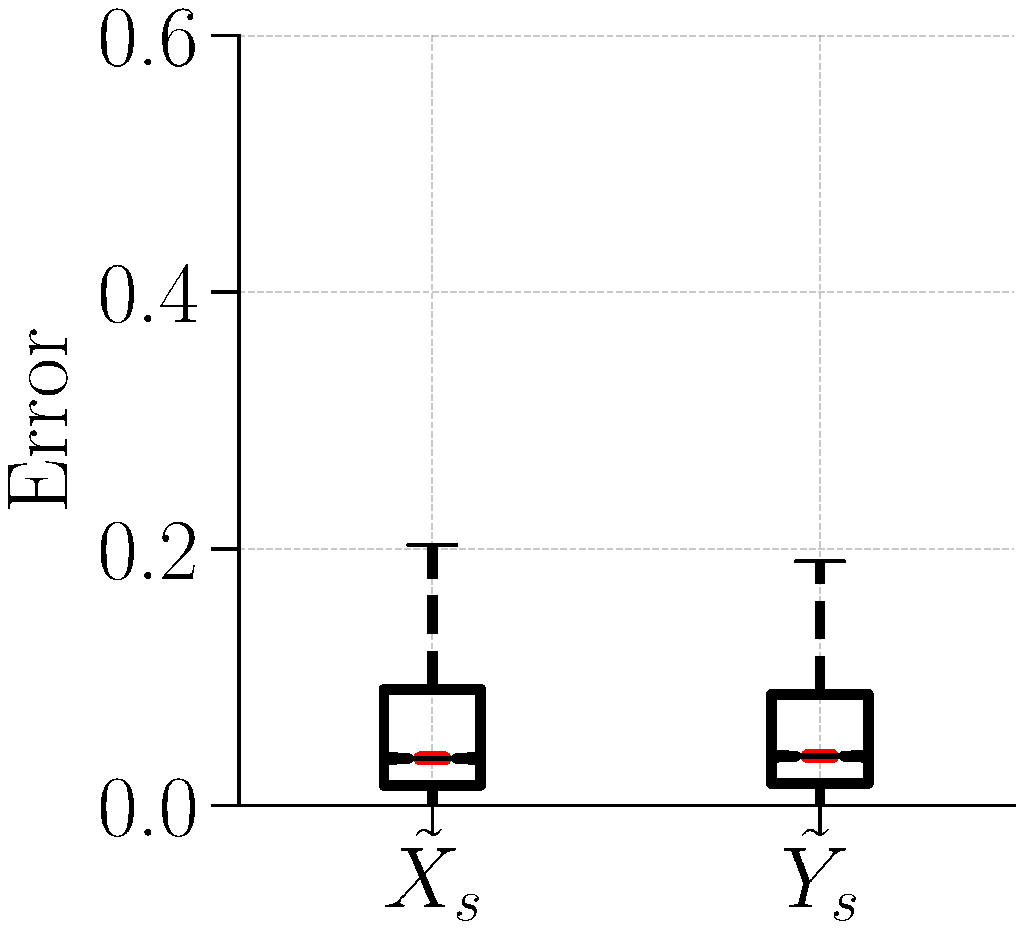}\label{box_plot_3point_10spacing}}
  \hspace{2mm} \subfloat[$(5,2)$-Sensor array] {\includegraphics[width=
    0.22\columnwidth]{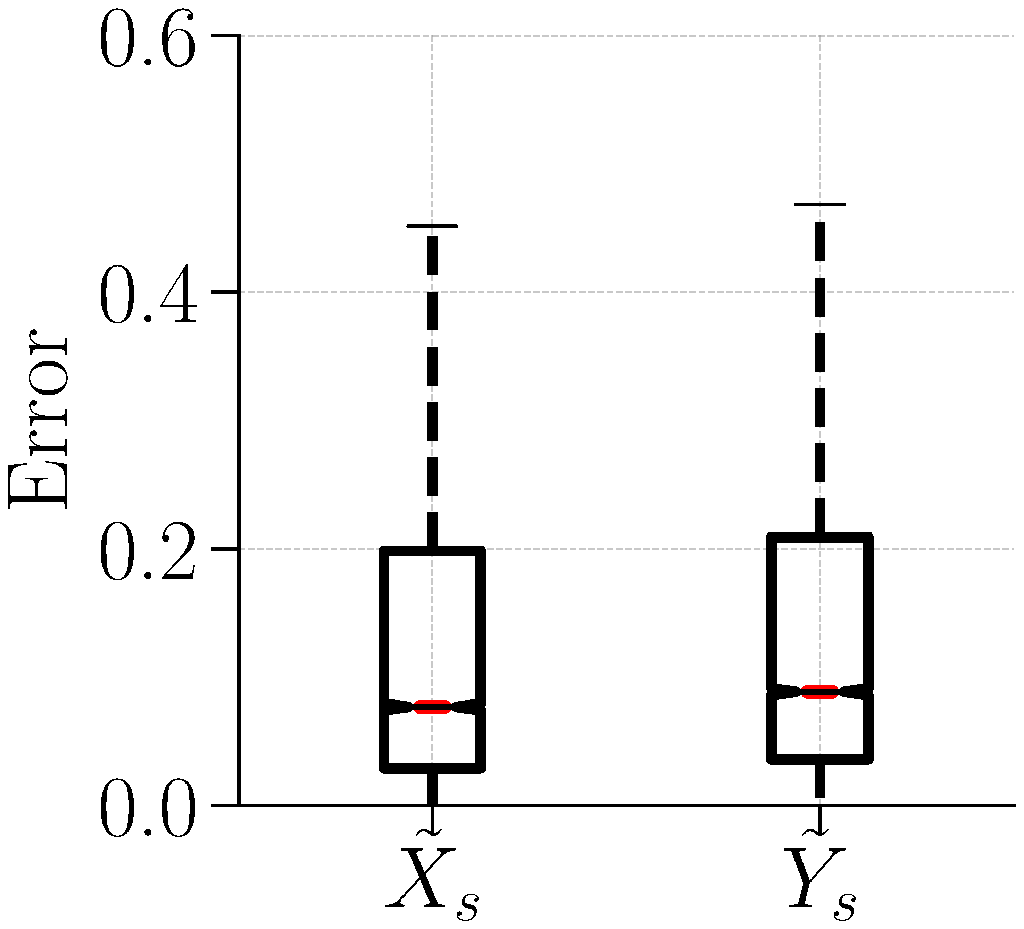}\label{box_plot_5point_2spacing}}
  \subfloat[$(5,5)$-Sensor array] {\includegraphics[width=
    0.22\columnwidth]{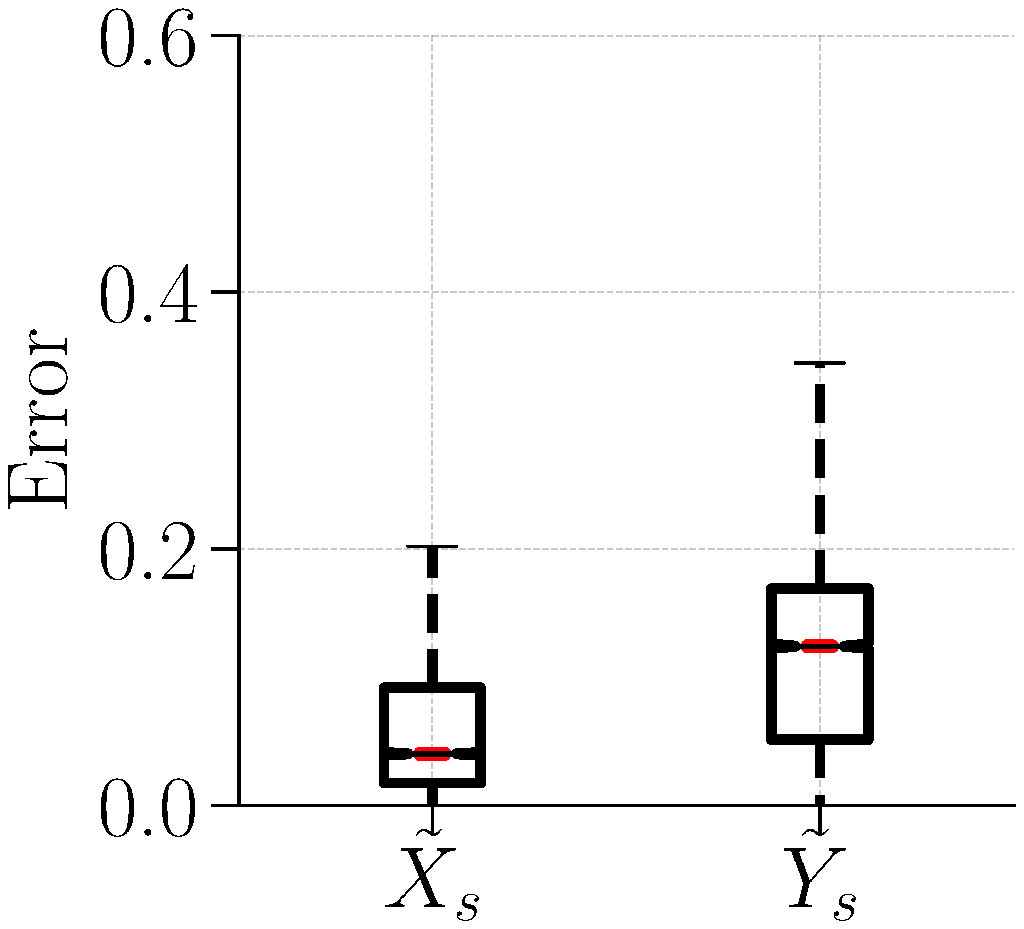}\label{box_plot_5point_5spacing}}
  \subfloat[$(5,10)$-Sensor array] {\includegraphics[width=
    0.22\columnwidth]{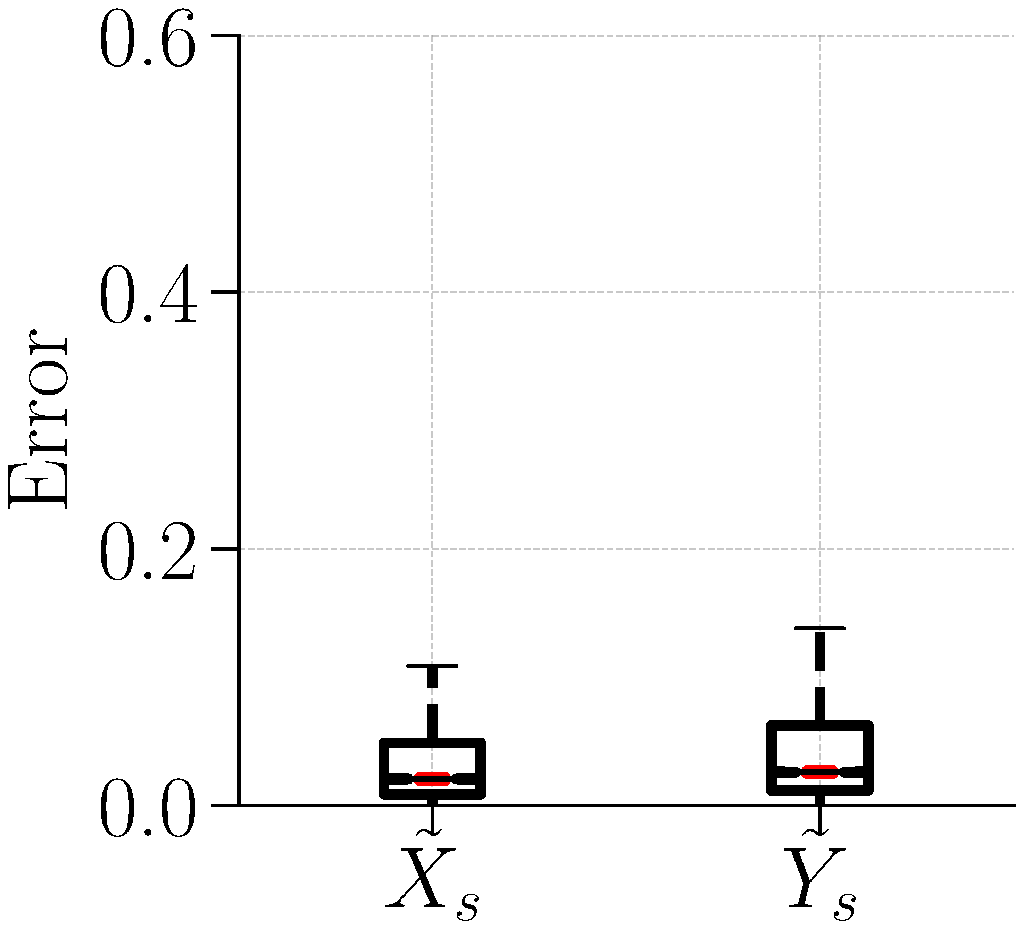}\label{box_plot_5point_10spacing}}
  \hspace{2mm} \subfloat[$(7,2)$-Sensor array] {\includegraphics[width=
    0.22\columnwidth]{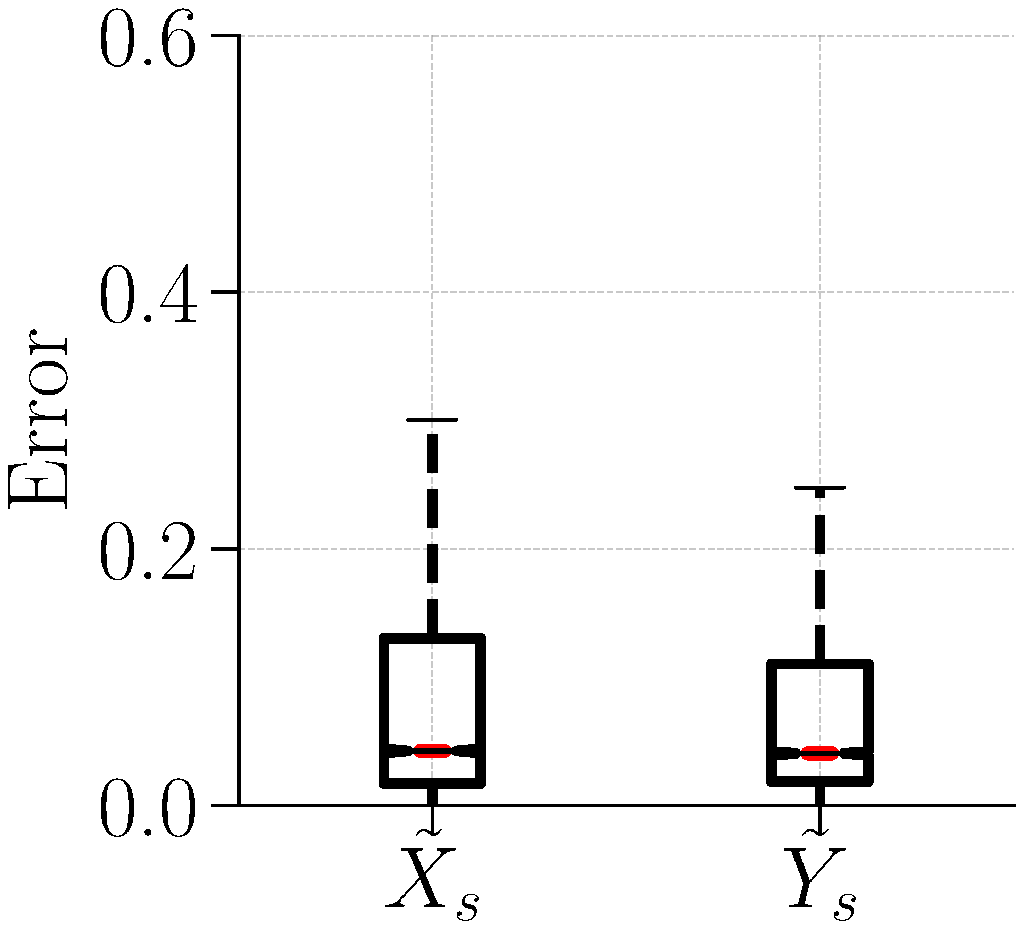}\label{box_plot_7point_2spacing}}
  \subfloat[$(7,5)$-Sensor array] {\includegraphics[width=
    0.22\columnwidth]{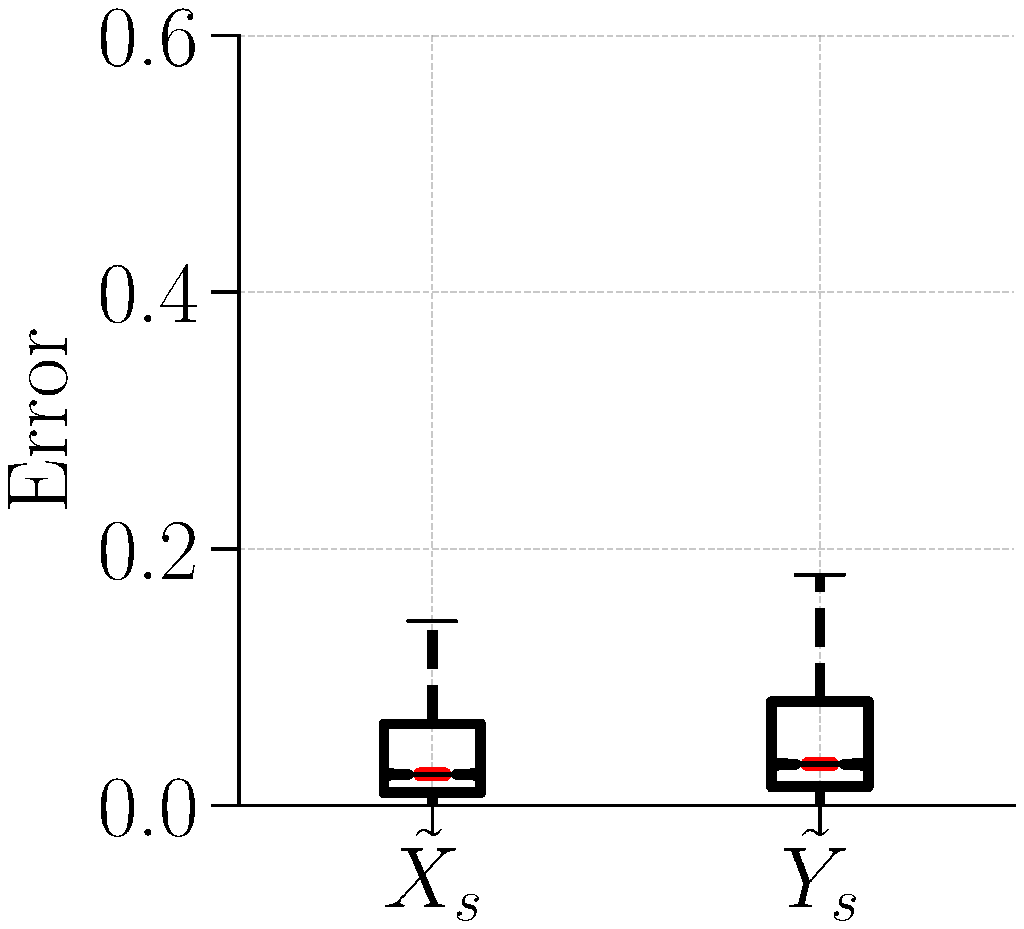}\label{box_plot_7point_5spacing}}
  \subfloat[$(7,10)$-Sensor array] {\includegraphics[width=
    0.22\columnwidth]{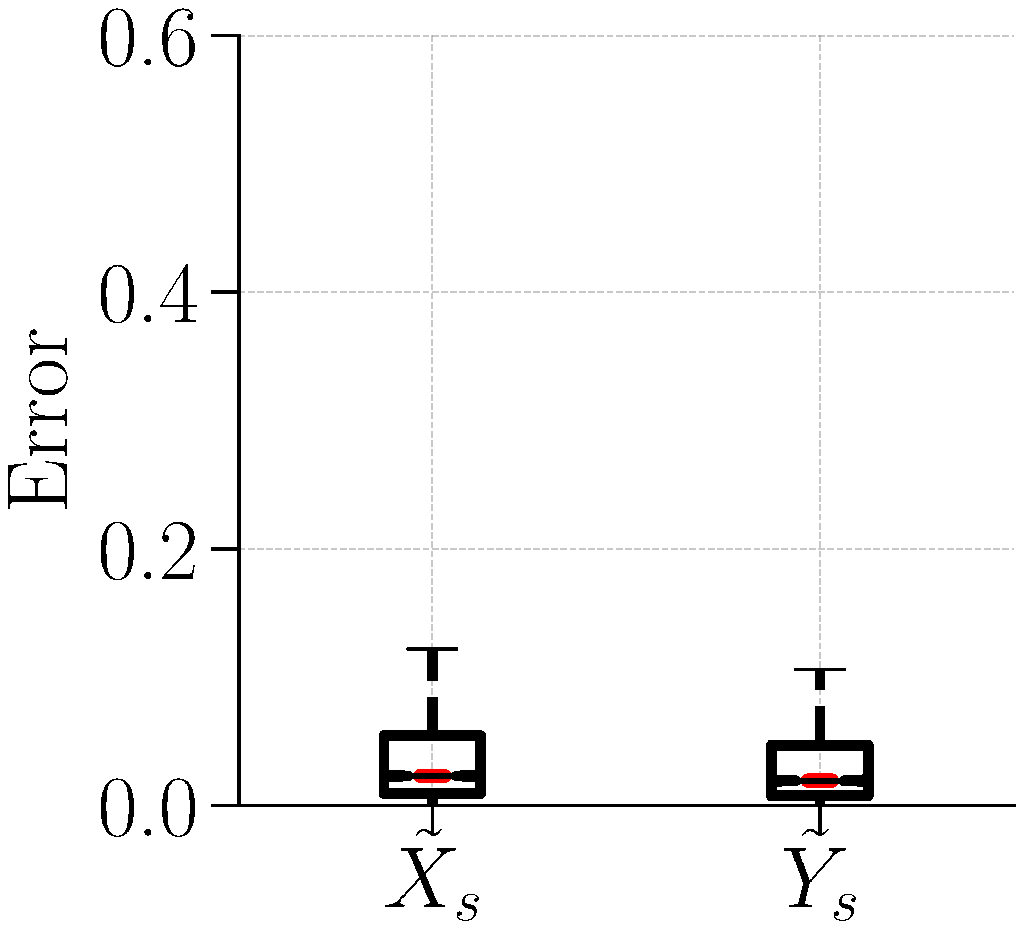}\label{box_plot_7point_10spacing}}
  \caption{Performance of the ANN-based regressor in estimating the location
    of the flow singularity $(X_s,Y_s)$. The distribution of the relative
    error is shown by means of boxplots representing the the median error in
    red, and box extremities corresponding to the 25\uth and 75\uth
    percentiles of the distribution of error, and whiskers being at the 10\uth
    and 90\uth percentiles.}
  \label{Location accuracy single source}
\end{figure}

\begin{figure}[htbp!]
  \centering \subfloat[Estimation of the doublet center location]
  {\includegraphics[width=
    0.25\columnwidth]{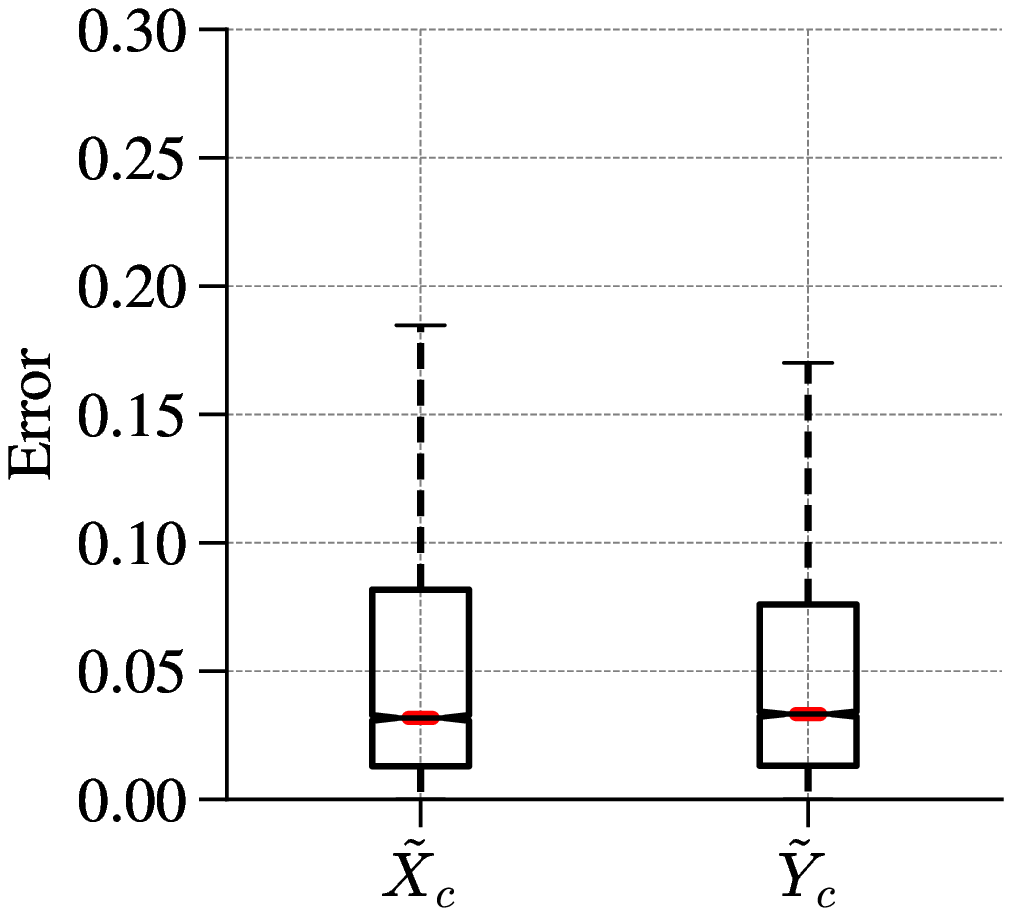}\label{center_location_dipole_constant_strength_res_vel_pi}}
  \subfloat[Estimation of the doublet orientation $\alpha$]
  {\includegraphics[width= 0.25\columnwidth]{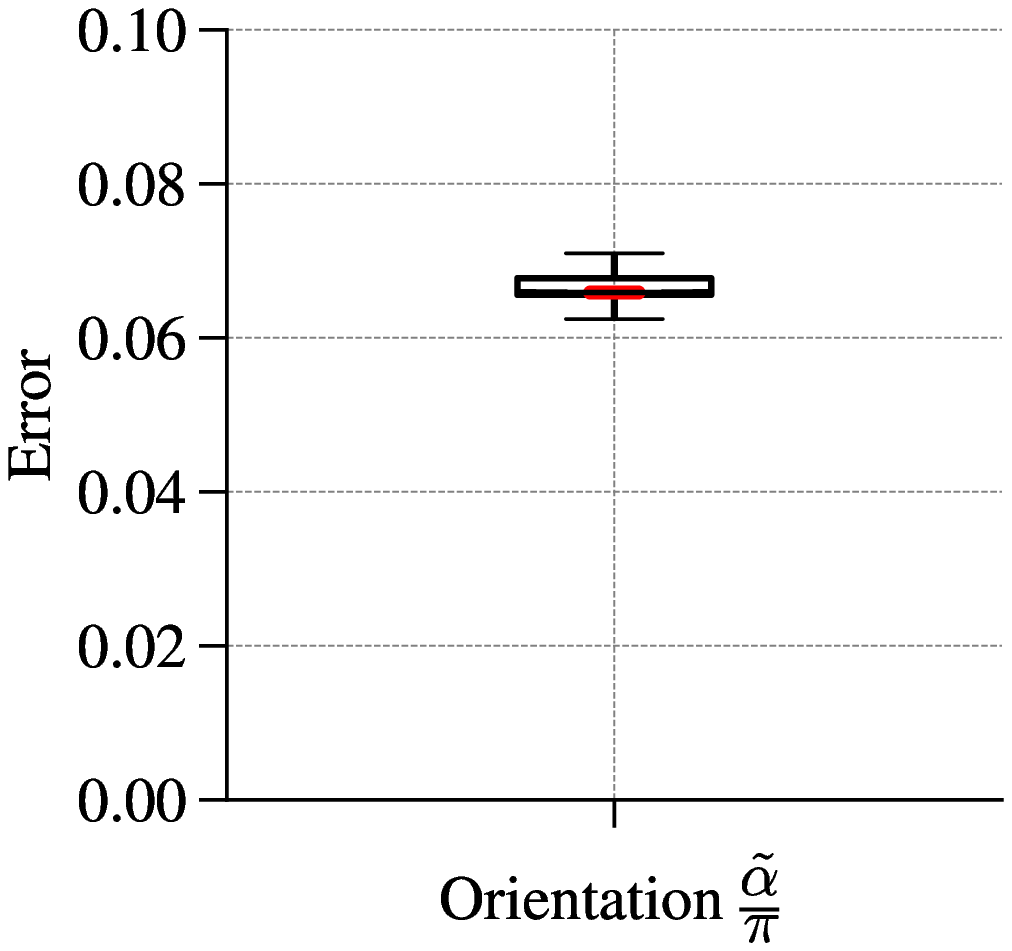}\label{orientation
      alpha estimation}}
  \caption{Performance of the ANN-based regressor in estimating the triplet
    $(\alpha,X_c,Y_c)$ characterizing doublet flows using a $(5,10)$-Sensor
    array. The distribution of the relative error is shown by means of
    boxplots representing the the median error in red, and box extremities
    corresponding to the 25\uth and 75\uth percentiles of the distribution of
    error, and whiskers being at the 10\uth and 90\uth percentiles.}
  \label{Error in estimation of center location and doublet orientation}
\end{figure}

Next, we use an ANN-based regressor to estimate the triplet $(\alpha,X_c,Y_c)$
characterizing the orientation and center of a doublet flow (see Methods). The
distribution of relative error is shown in Fig.~\ref{Error in estimation of
  center location and doublet orientation} when using the flow speed
$\sqrt{u(x,y)^2+v(x,y)^2}$ to generate the sensed data with the
$(5,10)$-sensor array located at a $(x_a = 2\ello, y_a =- 2\ello)$ from the
origin. This 3-parameter regression task being more demanding than the
identification of a single source/sink, we observe that the estimation of the
doublet center location $(X_c,Y_c)$ is slightly less accurate than in the
single source/sink localization case, with a broader distribution of errors
(see Fig.~\ref{center_location_dipole_constant_strength_res_vel_pi}). It is
also worth noticing that the estimation of the doublet orientation $\alpha$ is
noticeably more challenging than its localization (see Fig.~\ref{orientation
  alpha estimation}).

\subsection*{Object shape identification}

As a final step, we consider the object shape identification problem, which
requires estimating the triplet $(\mu_1,\mu_2,\mu_3)$ of shape coefficients
from distant hydrodynamic measurements (see Methods). An important question is
which hydrodynamic sensed data yields the best estimation of the triplet. As
already mentioned, in the natural world, the LLS constitutes an array of dual
mechanosensors (canal and superficial neuromasts) thereby giving access to
both pressure and velocity fluctuations.  We therefore consider the
effectiveness of the ANN-based regression with four different sensed data: (i)
the $x$-component $u$ of the velocity field, (ii) the $y$-component $v$, (iii)
the flow speed $\sqrt{u^2+v^2}$, and (iv) $(u^2+v^2)$, which essentially
amounts to the dynamic pressure. The median estimation errors for all four
cases are reported in Table~\ref{tab:Median Errors}.

\begin{table}[htbp]
  \centering
  % the \resizebox{\columnwidth}{!}{\begin{tabular}{ |c|c|c|c| }
  \begin{tabular}{ l||c|c|c }
    \textbf{Sensed Data}&\textbf{$\mu_1$} &\textbf{$\mu_2$} &\textbf{$\mu_3$}\\ \hline
    \hline
    $u$ & $0.0036$ & $0.0082$ & $0.0181$  \\ \hline
    $v$  &  $0.0051$ & $0.0172$ & $0.0219$  \\ \hline
    $\sqrt{u^2 +v^2}$ & $0.0056$ & $0.0127$ & $0.0178$  \\ \hline
    $u^2 +v^2$  & \mbox{ }$0.0040$\mbox{ } & \mbox{ }$0.0077$\mbox{ } & \mbox{ }$0.0125$\mbox{ }  \\ \hline
  \end{tabular}
  \caption{Median relative error in estimating the shape coefficients
    $(\mu_1,\mu_2,\mu_3)$ using different sensed data. The first column
    specifies the sensed data serving as model training data. The other three
    columns report the median relative error in estimating $\mu_1$,~$\mu_2$,
    and $\mu_3$. The results are obtained using a $(5,10)$-sensor array
    located at $(x_a = 2\ello, y_a =- 2\ello)$ measured from the origin.}  
  \label{tab:Median Errors}
\end{table}

Interestingly, when considering the dynamic pressure $(u^2 +v^2)$ as the
sensed data, the ANN-based regression yields a significant improvement over
all other 3 options, and that for all 3 shape coefficients. At this stage, it
is worth highlighting that for the sink/source identification case, $u$ has
been found to be the best option, while for the doublet flow identification,
the flow speed $\sqrt{u^2+v^2}$ led to the best results. These results are
further confirmed in Fig.~\ref{Error in estimation of object shape
  coefficients with varying parameters}, which shows the full distribution of
relative errors for all 4 options of sensed data. We are therefore led to
conclude that there is no absolute---i.e. problem independent---best choice
for the sensed data, which is consistent with the dual mechanosensory nature
of the LLS.

The other important element gathered from Table~\ref{tab:Median Errors} and
Fig.~\ref{Error in estimation of object shape coefficients with varying
  parameters} is the hierarchy in error when going from $\mu_1$, to $\mu_2$,
and ultimately to $\mu_3$. This observation is consistent with what has
previously been reported by Bouffanais et
al.~\cite{bouffanais2010hydrodynamic}, and highlights the inherently
progressive perceptual shape discrimination capability of hydrodynamic object
identification.

\begin{figure}[!htb]
  \centering \subfloat[$u$] {\includegraphics[width=
    0.25\columnwidth]{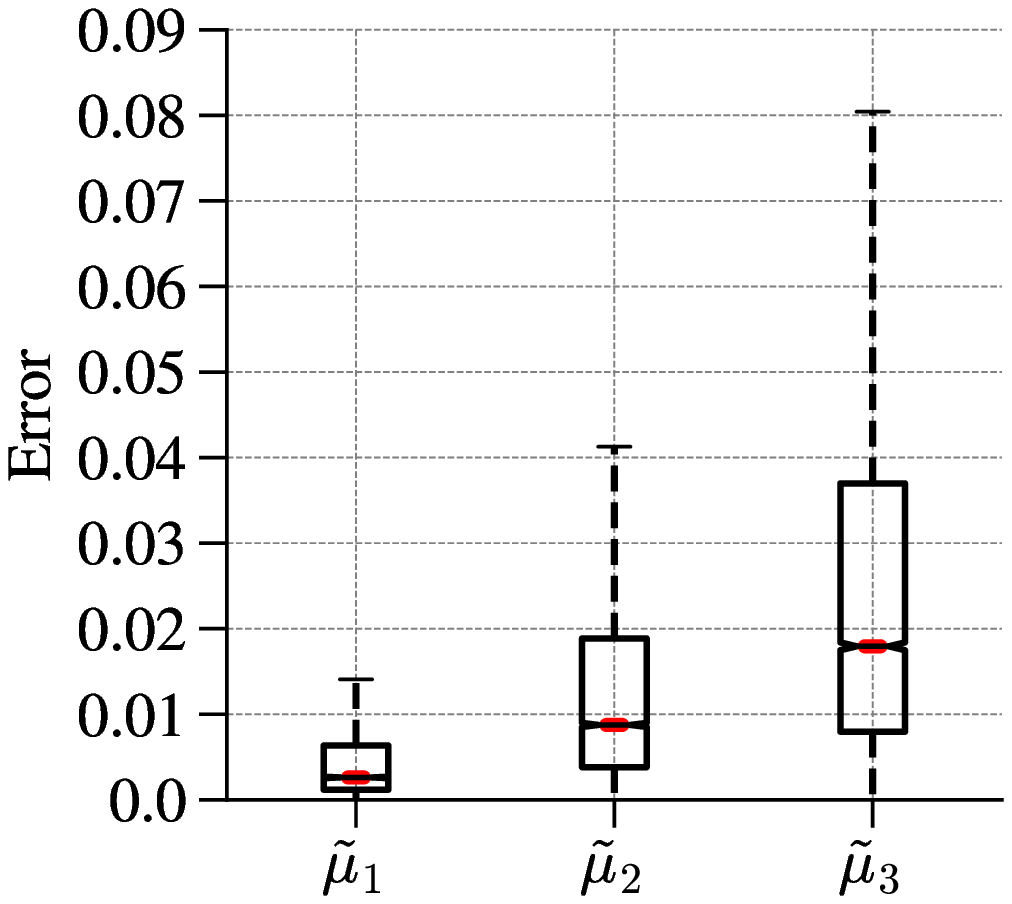}\label{object horz vel alpha}}
  \subfloat[$v$] {\includegraphics[width=
    0.25\columnwidth]{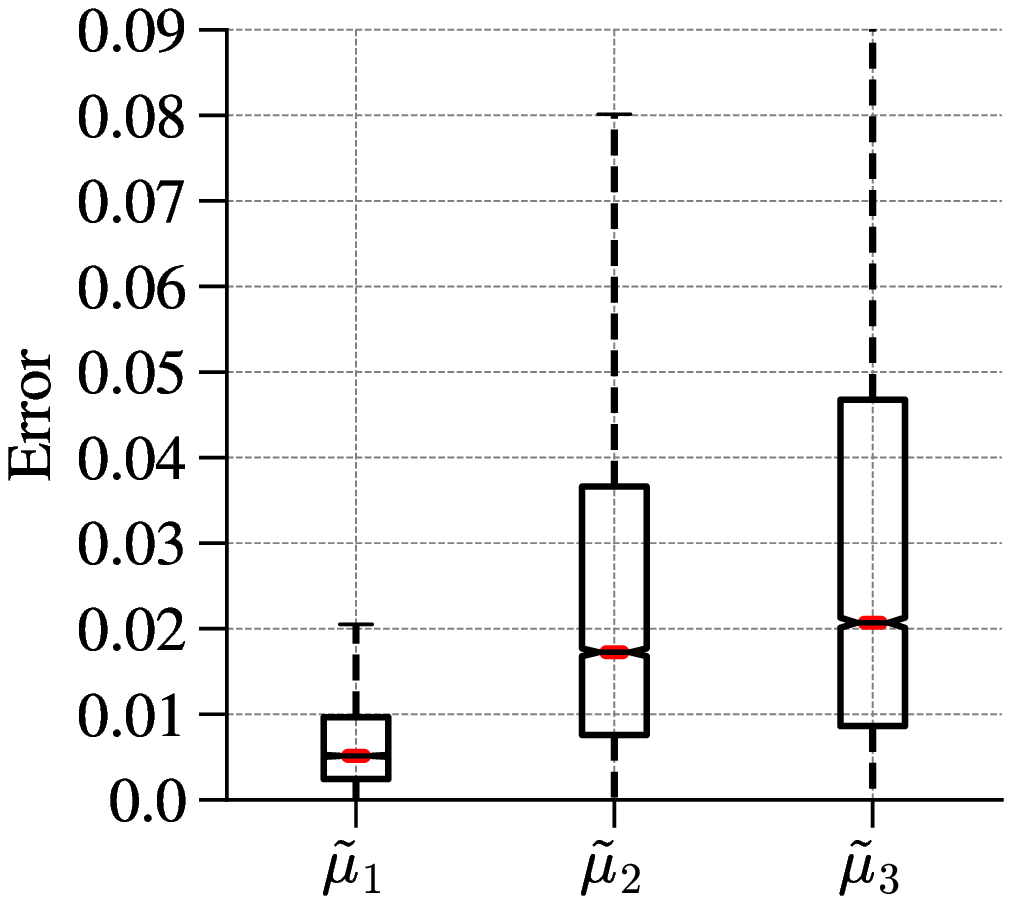}\label{object vert vel alpha}}
  \hspace{0mm} \subfloat[$\sqrt{u^2 +v^2}$]{\includegraphics[width=
    0.25\columnwidth]{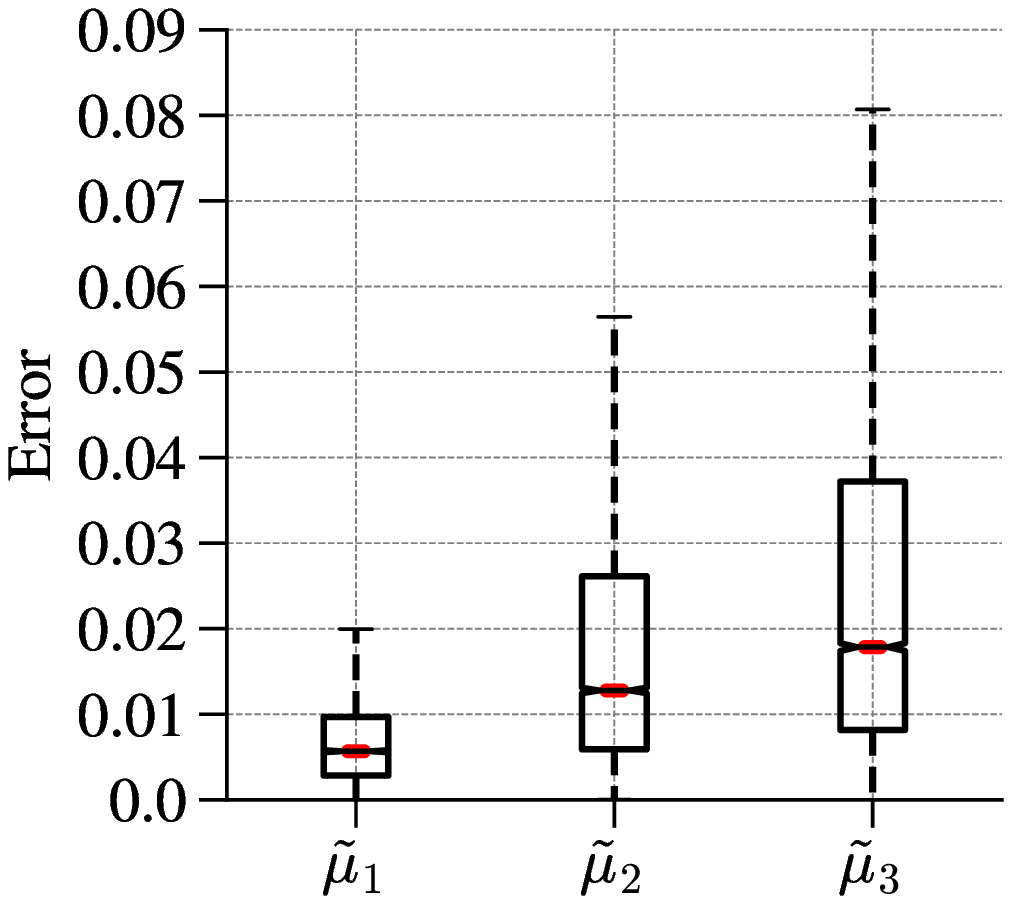}\label{object res vel alpha}}
  \subfloat[$u^2 +v^2$]{\includegraphics[width=
    0.25\columnwidth]{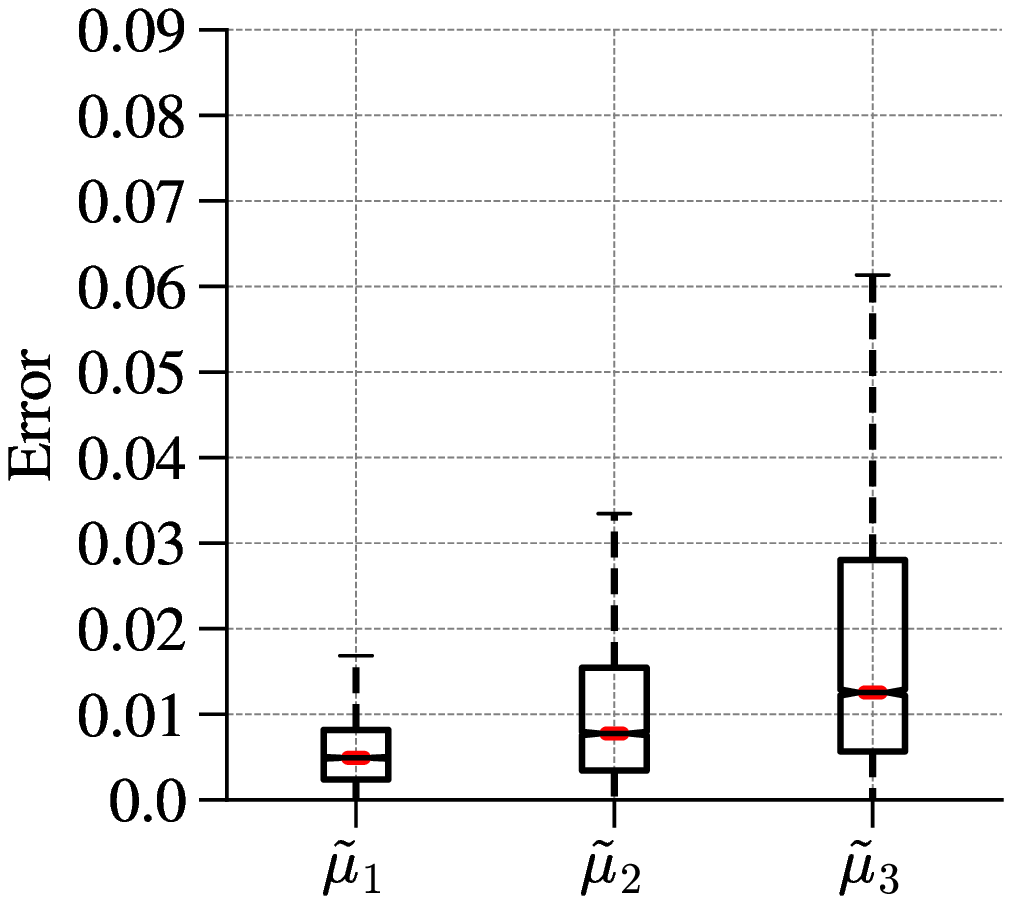}\label{object pressure alpha}}
  \caption{Distribution of relative error in estimating the shape coefficients
    for all four options of sensed data. The results are obtained using a
    $(5,10)$-sensor array located at $(x_a = 2\ello, y_a =- 2\ello)$ measured
    from the origin.}
  \label{Error in estimation of object shape coefficients with varying
    parameters}
\end{figure}

\begin{figure}[!htb]
  \centering \subfloat[]{\includegraphics[width=
    0.45\columnwidth]{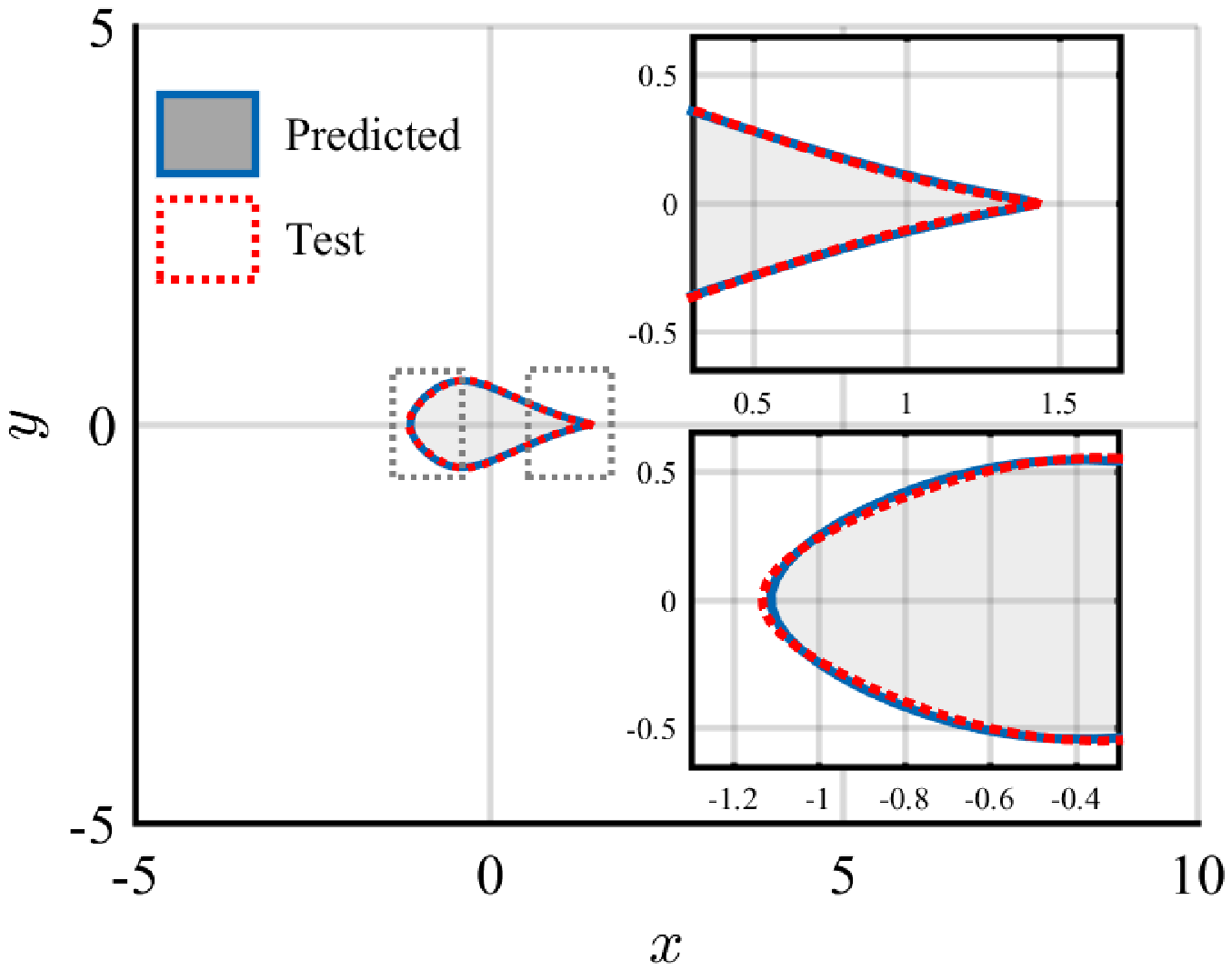}\label{object_2l0}} \hspace{0mm}
  \subfloat[]{\includegraphics[width=
    0.45\columnwidth]{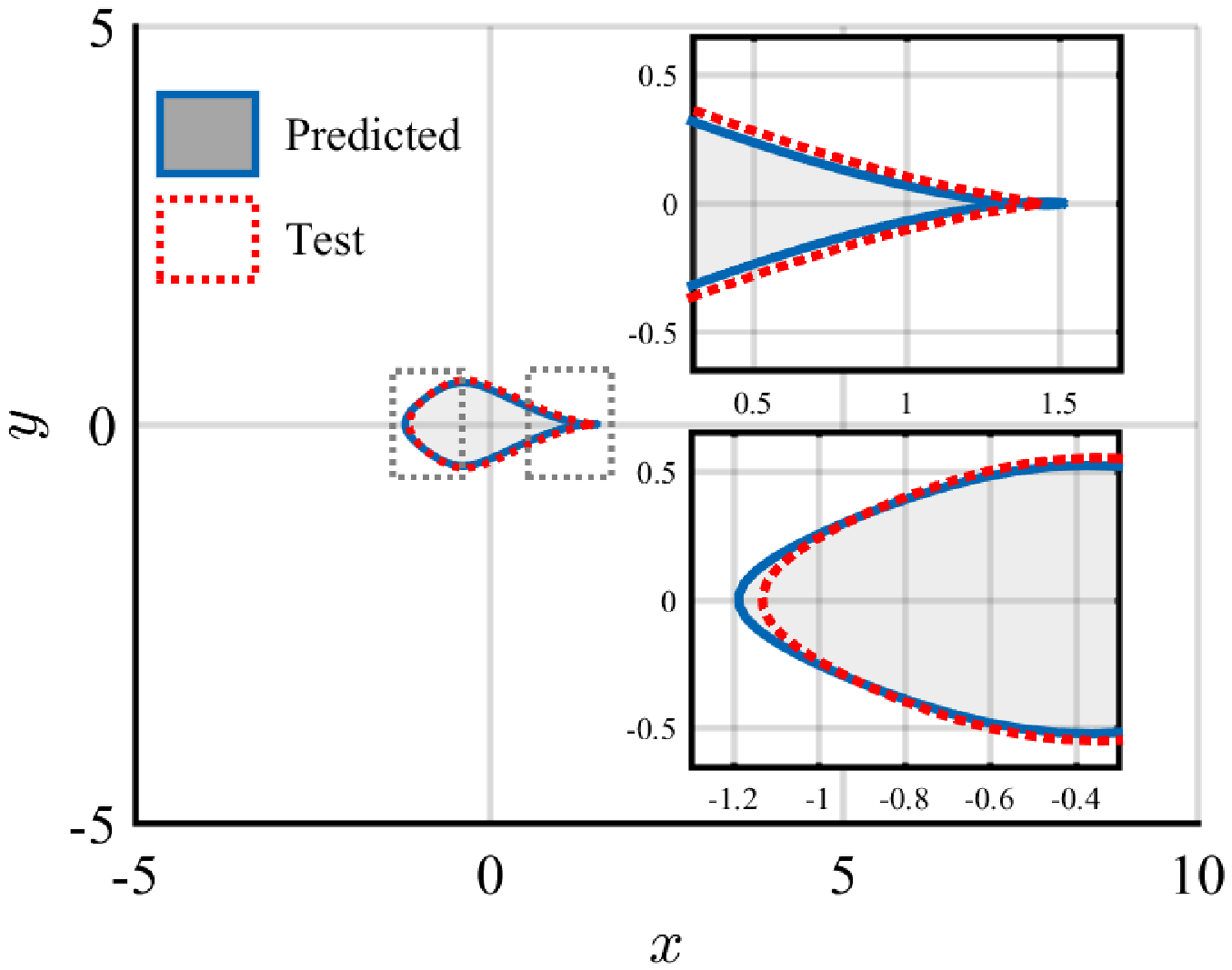}\label{object_9l0}}
  \caption{Direct visual comparison of the actual shape of the obstacle (test
    shape limited by the dotted line corresponding to $\mu_1=1/3$,
    $\mu_2=1/6$, and $\mu_3=1/12$) with the predicted shape (solid line) at
    two different distances: (a) sensor array in the near-field region for a
    distance $d=2.8\ello$ ($\tilde{\mu}_1 = 0.3363$, $\tilde{\mu}_2= 0.1737$,
    $\tilde{\mu}_3 = 0.0711$), and (b) sensor array in the far-field region
    for a distance $d=9.2\ello$ ($\tilde{\mu}_1 = 0.3517$, $\tilde{\mu}_2 =
    0.1902$, $\tilde{\mu}_3 = 0.1227$).}
  \label{object shape contour comparison}
\end{figure}

The final critical factor to be investigated is the influence of the distance
between the sensor array and the object whose shape is sought. Previous
attempts at detecting and identifying an obstacle by means of hydrodynamic
imaging were based on classical data processing approaches exhibiting severe
limitations in terms of their ability to extract shape features, even at short
distances away from the
obstacle~\cite{sichert2009hydrodynamic,bouffanais2010hydrodynamic}. Figure~\ref{Error
  vs distance from the object} shows the performance of the ANN-based
regression for all three shape coefficients at various positions behind the
obstacle (see some positions of the sensor array highlighted in red in
Fig.~\ref{fig:Computational Domain}). These results were obtained by sensing
the dynamic pressure $(u^2+v^2)$ with a $(5,10)$-sensor array located along
the line $y_a=-2\ello$ with $x_a$ varying from $2\ello$ to approximately
$10\ello$. On the horizontal axis of Fig.~\ref{Error vs distance from the
  object}, one finds the distance $d=\sqrt{x_a^2+y_a^2}$ between the top-left
corner of the array and the conformal center of the obstacle. As expected, the
relative error in estimating the shape coefficients $\{ \mu_k\}_{k=1,2,3}$
increases with the distance $d$. However, the performance of the ANN-based
regression is outstanding for $\mu_1$ with less than 5\% relative error at
very large distances away from the obstacle ($\sim 10\ello$). For the
triangular coefficient $\mu_2$, the performance is remarkably good (relative
error around 10\% at $5\ello$, and slightly above 15\% at $9.2\ello$) compared
to previously obtained
results~\cite{sichert2009hydrodynamic,bouffanais2010hydrodynamic}. As
anticipated, the estimation of the quadrangular coefficient $\mu_3$ is much
more challenging, and this quantity is only reasonably estimated for distances
between the sensor array and the obstacle below $4\ello$. However, it is worth
doing a direct visual comparison of the actual shape of the obstacle (test
shape) with the predicted shape, at two different distances, and that combines
the use of all three shape coefficients: (a) sensor array in the near-field
region for a distance $d=2.8\ello$, and (b) sensor array in the far-field
region for a distance $d=9.2\ello$. The obstacle shapes for both of these
cases are shown in Fig.~\ref{object shape contour comparison}. When sensing in
the near-field region (Fig.~\ref{object_2l0}), the difference between test
shape and predicted one is barely noticeable. Interestingly, when sensing in
the far-field region (Fig.~\ref{object_9l0}), although the median relative
error in $\mu_3$ is close to 40\%, the predicted shape is visually extremely
similar to the test shape. For most practical purposes, it may be concluded
that the predicted shape is sufficiently close to the test one.

\begin{figure}[!htb]
  \centering {\includegraphics[width= 0.45\textwidth]{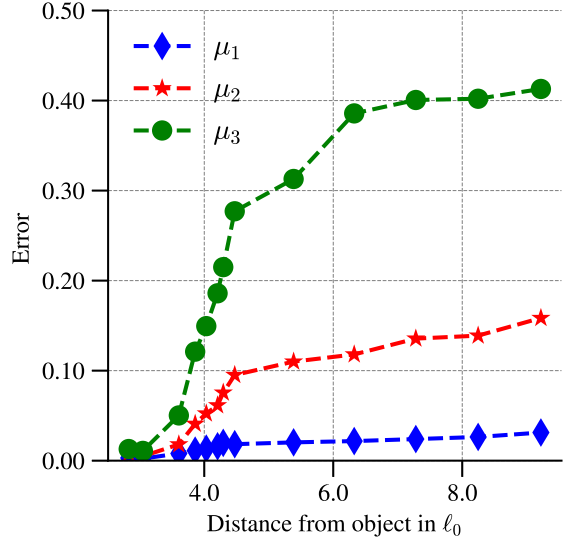}
    \caption{Median relative error in estimating all three shape coefficients
      when increasing the distance $d$ (measured in $\ello$ units) between the
      sensor array and the object.}
    \label{Error vs distance from the object}}
\end{figure}

\section*{Conclusion}

In this work, we considered the problem of hydrodynamic object identification
from remotely sensed flow data in the potential flow framework. We proposed
and implemented a neural data-processing model exhibiting vastly superior
performance compared to previously considered techniques (maximum likelihood
estimator and dynamic filtering). This approach uses artificial neural
networks that are trained with a large data set conveniently obtained from the
analytical solution to the forward problem associated with our inverse problem
of object shape identification.

The effectiveness of our neural networks regression and classification is
assessed on two singular potential flows: source/sink flow, and doublet
flow. Classical linear regression techniques are found to be completely
ineffective in identifying both singularities. The influence of the sensor
array design is analyzed, thereby revealing that a relatively small and static
array with 25 sensing units is sufficient for the considered task. The
ANN-based obstacle identification further confirms the progressive perceptual
character of this hydrodynamic shape discrimination capability. It also shows
remarkable performance even at relatively large distances away from the
obstacle.

Moreover, the ANN-based data-processing methodology reported here is being
further developed to tackle complex fluid flow problems involving unstable
swirling flows~\cite{A7}, as well as the identification of complex
relationships between flow variables in turbulent channel flows.

Finally, it is important stressing that the combination of our ANN-based
data-processing technique with recent hardware advances in MEMS paves the way
to the design and development of full-fledged artificial lateral-line systems
that could be integrated to the next generation of engineered underwater
vehicles and robots.

% ---------------------------------------------------------------------------------------------------
%\bibliography{Bibliography}

\section*{Acknowledgments}

This research was supported by a MOE-SUTD PhD fellowship (SL) and by SUTD
grants (IDC: IDG31800101 and AI Sector: PIE-SGP-AI-2018-01). We thank
Dr.~David Mateo for his assistance in data visualization.

\section*{Author contributions statement}

R.B. and S.L. designed the study. S.L. performed research, developed the
methods. B.B.T. and S.L. jointly developed the artificial neural
model. R.B. coordinated the study. S.L. and R.B. analyzed the results and
wrote the paper. All authors reviewed the manuscript.

\section*{Additional information}

\textbf{Supplementary information} accompanies this paper at doi:

\textbf{Competing Interests:} The authors declare that they have no competing interests.

\end{document}